\documentclass[11pt]{article}

\usepackage[left=2cm, right=2cm, top=1cm,bottom=2cm]{geometry}
\usepackage{natbib,url}
\usepackage{amsfonts}
\usepackage{graphicx, amsmath, float}
\usepackage{mathtools}
\usepackage{commath}
\usepackage[sc,osf]{mathpazo}
\usepackage{longtable}
\usepackage{setspace}
\usepackage[compact]{titlesec}

\usepackage{tikz,color}
\usetikzlibrary{arrows,calc,positioning}
\usetikzlibrary{arrows,shapes.arrows,calc,shapes.geometric,shapes.multipart,decorations.pathmorphing,positioning}

\usepackage[justification=centering]{caption}
\usepackage{subfigure}
\usepackage{amssymb}
\usepackage{mathrsfs}
\usepackage{verbatim}		

\usepackage{threeparttable}
\usepackage{multirow}
\makeatletter
\def\hlinewd#1{%
  \noalign{\ifnum0=`}\fi\hrule \@height #1 \futurelet
   \reserved@a\@xhline}
\makeatother

\newtheorem{thm}{Theorem}

\usepackage{enumerate}
\usepackage{appendix}

\usepackage{siunitx}

\allowdisplaybreaks

\title{Causal Inference for Comprehensive Cohort Studies \footnote{Initiation Submission to {\em Biometrics}:  October, 2016; Re-Submission to {\em Biometrics}: July, 2019}}
\author{Yi Lu$^{1}$,
Daniel O. Scharfstein$^{1,*}$, 
Maria M. Brooks$^2$, \\Kevin Quach$^{1}$ and Edward H. Kennedy$^{3}$\\
$^1$Johns Hopkins Bloomberg School of Public Health, Baltimore, MD, U.S.A.\\
$^2$University of Pittsburgh Graduate School of Public Health, Pittsburgh, PA, U.S.A. \\
$^3$ Carnegie Mellon University, Pittsburgh, PA, U.S.A.}
\date{\today}

\begin{document}

\maketitle

\label{firstpage}

\begin{abstract}
In a comprehensive cohort study of two competing treatments (say, A and B), clinically eligible individuals are first asked to enroll in a randomized trial and, if they refuse, are then asked to enroll in a parallel observational study in which they can choose treatment according to their own preference. We consider estimation of two estimands:  (1) comprehensive cohort causal effect -- the difference in mean potential outcomes had all patients in the comprehensive cohort received treatment A vs. treatment B and (2) randomized trial causal effect -- the difference in mean potential outcomes had all patients enrolled in the randomized trial received treatment A vs. treatment B. For each estimand, we consider 
inference under various sets of unconfoundedness assumptions and construct semiparametric efficient and robust estimators.  These estimators depend on nuisance functions, which we estimate, for illustrative purposes, using generalized additive models.  Using the theory of sample splitting, we establish the asymptotic properties of our proposed estimators.  We also illustrate our methodology using data from the Bypass Angioplasty Revascularization Investigation (BARI) randomized trial and observational registry to evaluate the effect of percutaneous transluminal coronary balloon angioplasty versus coronary artery bypass grafting on 5-year mortality.  To evaluate the finite sample performance of our estimators, we use the BARI dataset as the basis of a realistic simulation study.
\end{abstract}


\newpage

\section{Introduction}

Randomized controlled trials (RCTs) are considered to be the gold standard for comparing treatments, primarily because the experimental design probabilistically ensures that treatment groups are balanced with respect to measured and unmeasured prognostic factors.  A well conducted RCT is said to have high internal validity.  However, its external validity (i.e., generalizability of results to a broader population) is not guaranteed.  This is because eligible patients who agree to enroll in an RCT may 
be not be a representative sample of all eligible patients.  Due to the tension between internal and external validity, researchers have recommended that all clinically eligible patients, agreeing to randomization or not, should be enrolled and studied \citep{fielding1999patients}. 

\cite{olschewski1985CCS} introduced the comprehensive cohort study (CCS) design for evaluating competing treatments (say, A and B) in which clinically eligible participants are first asked to enroll in a randomized trial and, if they refuse, are then asked to enroll in a parallel observational study (OBS) in which they can choose treatment according to their own preference. 


Most of the literature that describes methods for analyzing data from a CCS has focused on estimating treatment effects separately for the RCT and OBS (e.g., \cite{olschewski1992analysis}, \cite{henshaw1993comparison}, \cite{nicolaides1994comparison}, \cite{schmoor1996randomized},  \cite{king1997angioplasty}, \cite{detre1999coronary}, \cite{bedi2000assessing}, \cite{brooks2000predictors}, \cite{kerry2000routine}, \cite{king2000randomised}, \cite{rovers2001generalizability}, \cite{jensen2003hormone}, \cite{schmoor2008evidence}), with some adjustment for confounding in the OBS.  With the exception of \cite{olschewski1992analysis}, \cite{king1997angioplasty} and \cite{brooks2000predictors}, there is no borrowing of information between the RCT and OBS. 

Here, we focus on drawing causal inferences about treatment effects from a CCS, where the primary outcome (continuous or binary) is to be measured at a fixed point in time after treatment assignment.  We are interested in drawing inference about two causal estimands:   {\em comprehensive cohort causal effect} -- the difference in mean potential outcomes had all patients in the CCS received treatment A vs. treatment B, and {\em randomized trial causal effect} -- the difference in mean potential outcomes had all patients enrolled in the RCT received treatment A vs. treatment B.   
The comprehensive cohort causal effect is of interest because it refers to a broader population of eligible patients than those willing to participate in the randomized trial. The randomized trial causal effect is of interest because it is defined experimentally and its estimates would more confidently generalize to future patients with covariate profiles similar to those in the RCT.   


The paper is organized as follows.  In Section \ref{chap2:sec:notation}, we introduce notation, data structure, and main assumptions. 
Sections \ref{cc} and \ref{rct} present estimators and their properties for the comprehensive cohort causal effect and randomized trial causal effect, respectively.
 In Section \ref{chap2:sec:data},  we illustrate our methods  using data from the Bypass Angioplasty Revascularization Investigation (BARI) randomized trial and observational registry to evaluate the effect of percutaneous transluminal coronary balloon angioplasty (PTCA) versus coronary artery bypass grafting (CABG) on 5-year mortality. 
Section \ref{chap2:sec:simu} presents a simulation study, motivated by the BARI study, that evaluates the performance of our estimators.  The last section is devoted to a discussion.

\section{Notation and framework} \label{chap2:sec:notation}

Let $X$ denote a vector of baseline covariates and let $Y$ be the observed outcome (continuous or binary).  Let $R$ denote the randomization consent indicator (1 for RCT, 0 for OBS) and let $T$ denote the treatment assignment indicator (1 for treatment A, 0 for treatment B).  The observed data for an individual are $O=(X',Y,R,T)'$. We assume $n$ independent and identically distributed copies of $O$ are drawn from some distribution $P^*$ contained in ${\cal M}$.  Throughout, the superscript $*$ will be used to denote the true value of the quantity to which it is appended.  The subscript $i$ will denote data associated with individual $i$.

Let $Y_{1,r}$ and $Y_{0,r}$ be an eligible patient's potential outcome under treatment $A$ and $B$, respectively, when enrolled into study $r$ ($r=1$ for RCT and $r=0$ for OBS).  We  assume there is a \emph{single version of treatment} (i.e., $Y_{t,1}=Y_{t,0}=Y_t$ for $t=0,1$; \cite{vanderweele2009concerning}).  


Letting $\nu_t^* = E[Y_t|R=1]$, the randomized trial causal effect is defined as $\Delta_{RCT} = \nu_1^* - \nu_0^*$.  Letting $\mu_{t}^* = E[Y_t]$, the comprehensive causal effect is defined as $\Delta_{CC} = \mu_1^* - \mu_0^*$.  Our goal is to draw inference about $\Delta_{RCT}$ and $\Delta_{CC}$. 

To identify these causal effects from the observed data, we posit assumptions sufficient for identification of $\nu_t^*$ and $\mu^*_{t}$ ($t=0,1$).  We make the \emph{no interference assumption} so that the potential outcomes of an individual are unaffected by the randomization consent and treatment decision of any other individual \citep{cox1958planning}.  We make the \emph{consistency} assumption that connects the observed outcomes to the potential outcomes via the following relation: $Y = T Y_1 + (1-T) Y_0$ \citep{vanderweele2009concerning}.  In addition, we will utilize assumptions from among the following:
\begin{enumerate}[({A}1)]
\item \emph{In the RCT, treatment is randomized}: $\left. {T \bot \left( {{Y_1},{Y_0},X} \right)} \right|R = 1$
\item \emph{In the OBS, treatment is randomized within levels of $X$}: $\left. {T \bot \left( {{Y_1},{Y_0}} \right)} \right|R = 0,X$ 
\item \emph{Consent into the RCT is randomized within levels of $X$}:  $\left. {R \bot \left( {{Y_1},{Y_0}} \right)} \right|X$
\end{enumerate}
(A1) and (A2) indicate that, conditional on $(X,Y_1,Y_0, R)$, treatment selection is a Bernoulli process with probability $\pi_t^*(r,x) = P\left[ T = t |R=r,X=x \right]$. Note that (A1) implies that $\pi_t^*(1,x)$ does not depend on $x$. Letting $\pi_{t1}^* = \pi_t^*(1,x)$ be the {\em known} randomization probability, we have that $\pi_t^*(r,x) = r \pi_{t1}^* + (1-r) \pi_t^*(0,x)$. 
We let $\lambda_1^*(x) = P[R=1|X=x]$.  We further impose \emph{positivity} conditions so $0 < \lambda_1^*(x) < 1$ and $0 < \pi_1^*(r,x) <1$ for all $x$ and $r=0,1$ \citep{hernan2006estimating}.   We assume the first and second moments of the conditional distribution of $Y_t$ given $R=r$ and $X=x$ are finite for all $x$ and $r,t=0,1$. We define $\tau^*_t(r,x) = E[Y_t|R=r,X=x]$ and $\tau^*_t(x) = E[Y_t|X=x]$.  Note that under (A1) and (A2), $\tau_t^*(r,x) = E[Y|T=t,R=r,X=x]$. Under (A3), $\tau^*_t(r,x) = \tau^*_t(x)$.  For convenience,  Table \ref{chap2:tb:notation} provides a complete list of our notation.

\begin{table}
\centering
\caption{Notation} \label{chap2:tb:notation}
\begin{spacing}{1.2}
\begin{tabular}{l@{\hskip 0.3in}l}
  \hlinewd{1pt}
Symbol & Description ($t=0,1$) \\
  \hline
$R$ & Randomization consent indicator (1 for RCT, 0 for OBS)\\
$T$ & Treatment indicator (1 for A, 0 for B) \\
$X$ & Baseline covariates  \\  
$Y_t$ & Potential outcome under treatment $t$ \\
$Y$ & Observed outcome \\
$\mu_{t}^*$ & $E[Y_t]$ \\
$\Delta_{CC}$ & $\mu_{1}^* - \mu_{0}^*$ \\
$\nu_{t}^*$ & $E[Y_t | R=1]$ \\
$\Delta_{RCT}$ & $\nu_{1}^* - \nu_{0}^*$ \\
$\tau_{t}^*(r,x)$ & $E[Y_t |R=r, X=x] \stackrel{(A1,A2)}{=} E[Y|T=t,R=r,X=x]$ \\
$\tau_{t}^*(x)$ & $E[Y_t |X=x] \stackrel{(A3)}{=} \tau_{t}^*(r,x)  \stackrel{(A1,A2,A3)}{=} E[Y|T=t,X=x]$ \\
$\Delta^*_{CC}$ & $\mu_{1}^* - \mu_{0}^*$ \\
$\Delta^*_{RCT}$ & $\nu_{1}^* - \nu_{0}^*$ \\
$\lambda_r^*$ & $P[R=r]$ \\
$\lambda_r^*(x)$ & $P[R=r | X=x]$ \\
$\pi_t ^* (r,x)$ & $P[T=t|R=r,X=x]$ \\
$\pi_t ^* (x)$ & $P[T=t|X=x] = \pi_t(1,x) \lambda_1^*(x) + \pi_t(0,x) \lambda_0^*(x)$ \\
$\pi _{t1}^*$ & $P\left[ {T = t|R = 1} \right] \stackrel{(A1)}{=} \pi_t ^* (1,x)$ \\
   \hlinewd{1pt}
\end{tabular}
\end{spacing}
\end{table}

We consider inference about the comprehensive cohort causal effect under (A1) and (A2),  under (A1) and (A3), and under (A1), (A2) and (A3), and inference about the randomized trial causal effect under (A1) and under (A1), (A2) and (A3).    The two parameters will be equal if the causal effect in the OBS is equal to the causal effect in the RCT.  This will occur when (A3) holds and either (i) the distribution of covariates is the same in the OBS and the RCT or (ii) there is no treatment effect heterogeneity. 
In this paper, we derive, using semiparametric theory \citep{tsiatis2006semiparametric,van2003unified,bickel1993efficient}, efficient and robust estimators of these causal effects under these sets of assumptions.  

\cite{marcus1997assessing} discussed assumptions similar to (A1) to (A3) for identifying the comprehensive cohort causal effect.  The author discussed how unbiased estimates of the comprehensive cohort causal effect can be obtained using outcome/treatment/covariate data from RCT only, OBS only as well as the entire CCS.  Our proposed estimator of the comprehensive causal effect under (A1) and (A2) and under (A1), (A2) and (A3) uses outcome/treatment/covariate data from the entire CCS and, under (A1) and (A3), uses outcome/treatment/covariate data from just the RCT.  

Estimation of the randomized trial causal effect under (A1), (A2) and (A3) uses outcome/treatment/covariate data from the entire CCS.  Conceptually, one can think of our approach as using data from the OBS to emulate the RCT \citep{hernan2008observational} and combining the resulting estimate with the estimate that uses RCT data only.  If (A2) and (A3) are correct, we will have a more precise estimate of the randomized trial causal effect, but if the assumptions are wrong then the estimate may be biased.  This bias-variance tradeoff may be appropriate in the setting of under-powered randomized trials.  

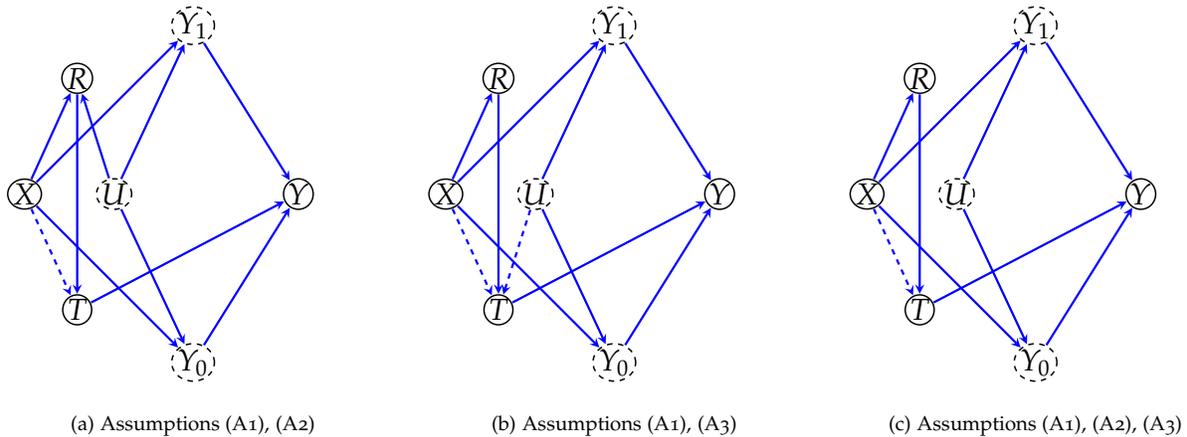
\begin{figure}[t]
	\begin{center}
		\scalebox{0.7}{
			\begin{tikzpicture}[>=stealth, node distance=1.2cm]
			\tikzstyle{format} = [draw, thick, circle, minimum size=1.0mm, inner sep=0pt]
			\tikzstyle{square} = [draw, thick, minimum size=1.0mm, inner sep=3pt]	
			
			\begin{scope}[xshift=0cm]
	
			\path[->, very thick]

			node [format, shape=ellipse, scale=1.5] (x) {$X$}
			node [format, shape=ellipse, above of=x, xshift=1cm,yshift=1cm, scale=1.5] (r) {$R$}
			node [format, shape=ellipse, below of=x, xshift=1cm,yshift=-1cm, scale=1.5] (t) {$T$}	
			node [format, shape=ellipse, right of=r, xshift=1cm,yshift=1cm, scale=1.5, dashed] (y1) {$Y_1$}	
			node [format, shape=ellipse, right of=t, xshift=1cm,yshift=-1cm, scale=1.5, dashed] (y0) {$Y_0$}	
			node [format, shape=ellipse, right of=x, xshift=4cm, scale=1.5] (y) {$Y$}	
			node [format, shape=ellipse, right of=x, xshift=0.5cm, scale=1.5,dashed] (u) {$U$}	

			(x) edge[blue] (r)
			(x) edge[blue,dashed] (t)
			(x) edge[blue] (y1)
			(x) edge[blue] (y0)
			(y1) edge[blue] (y)
			(y0) edge[blue] (y)
			(t) edge[blue] (y)
			(u) edge[blue] (y1)
			(u) edge[blue] (y0)
			(u) edge[blue] (r)
			(r) edge[blue] (t)
			node [below of=y0, xshift=0.0cm, yshift=0.0cm] (l) {(a) Assumptions (A1), (A2)}
			
			;
			
			\end{scope}

			\begin{scope}[xshift=8cm]
	
			\path[->, very thick]

			node [format, shape=ellipse, scale=1.5] (x) {$X$}
			node [format, shape=ellipse, above of=x, xshift=1cm,yshift=1cm, scale=1.5] (r) {$R$}
			node [format, shape=ellipse, below of=x, xshift=1cm,yshift=-1cm, scale=1.5] (t) {$T$}	
			node [format, shape=ellipse, right of=r, xshift=1cm,yshift=1cm, scale=1.5, dashed] (y1) {$Y_1$}	
			node [format, shape=ellipse, right of=t, xshift=1cm,yshift=-1cm, scale=1.5,dashed] (y0) {$Y_0$}	
			node [format, shape=ellipse, right of=x, xshift=4cm, scale=1.5] (y) {$Y$}	
			node [format, shape=ellipse, right of=x, xshift=0.5cm, scale=1.5, dashed] (u) {$U$}	

			(x) edge[blue] (r)
			(x) edge[blue,dashed] (t)
			(x) edge[blue] (y1)
			(x) edge[blue] (y0)
			(y1) edge[blue] (y)
			(y0) edge[blue] (y)
			(t) edge[blue] (y)
			(u) edge[blue] (y1)
			(u) edge[blue] (y0)
			(r) edge[blue] (t)
			(u) edge[blue,dashed] (t)
			node [below of=y0, xshift=0.0cm, yshift=0.0cm] (l) {(b) Assumptions (A1), (A3)}
			
			; 
			
			\end{scope}

			\begin{scope}[xshift=16cm]
	
			\path[->, very thick]

			node [format, shape=ellipse, scale=1.5] (x) {$X$}
			node [format, shape=ellipse, above of=x, xshift=1cm,yshift=1cm, scale=1.5] (r) {$R$}
			node [format, shape=ellipse, below of=x, xshift=1cm,yshift=-1cm, scale=1.5] (t) {$T$}	
			node [format, shape=ellipse, right of=r, xshift=1cm,yshift=1cm, scale=1.5, dashed] (y1) {$Y_1$}	
			node [format, shape=ellipse, right of=t, xshift=1cm,yshift=-1cm, scale=1.5, dashed] (y0) {$Y_0$}	
			node [format, shape=ellipse, right of=x, xshift=4cm, scale=1.5] (y) {$Y$}	
			node [format, shape=ellipse, right of=x, xshift=0.5cm, scale=1.5,dashed] (u) {$U$}	

			(x) edge[blue] (r)
			(x) edge[blue,dashed] (t)
			(x) edge[blue] (y1)
			(x) edge[blue] (y0)
			(y1) edge[blue] (y)
			(y0) edge[blue] (y)
			(t) edge[blue] (y)
			(u) edge[blue] (y1)
			(u) edge[blue] (y0)
			(r) edge[blue] (t)
			node [below of=y0, xshift=0.0cm, yshift=0.0cm] (l) {(c) Assumptions (A1), (A2), (A3)}
			
			;
			
			\end{scope}
			
			\end{tikzpicture}
		}
	\end{center}
	\caption{Directed acyclic graphs representing various assumptions. The dashed arrows in the figures indicate absence when $R=1$ and presence when $R=0$. Dashed nodes represent random variables that are unobserved ($U$) or potential outcomes ($Y_1,Y_0$).}
	\label{fig:dag}
\end{figure}

Figure \ref{fig:dag} presents directed acyclic graphs (DAGs) that represents various combinations of assumptions. The dashed arrows in the figures indicate absence when $R=1$ and presence when $R=0$.  Importantly, DAG (c), which encodes (A1), (A2) and (A3), induces testable restrictions, above and beyond (A1), on the distribution of the observed data; it implies that $Y$ is independent of $R$ given $T$ and $X$.  This holds because all paths from $R$ to $Y$ are blocked by $T$ and $X$, which are non-colliders.   Assuming (A1) holds, (A2) and (A3) can be tested by checking whether there is difference between the distribution of the observed outcome between those enrolled in the RCT and those enrolled in the OBS, after adjusting for treatment and covariates.  If  there is evidence of a difference, we cannot, unfortunately, tease out which of the two assumptions is misspecified.  Note that the conditional independence statement does not hold in DAG (a) because of the unblocked paths $R  \leftarrow U \rightarrow Y_t \rightarrow Y$, $t=0,1$, and in DAG (b) because of the paths $R  \rightarrow T \leftarrow U \rightarrow Y_t \rightarrow Y$, $t=0,1$, with the conditioning variable $T$ being a collider.

\section{Inference} \label{inference}

\subsection{Semiparametric Theory and Inferential Strategy}

Under assumptions, the target parameters $\mu_t^*$ and $\nu_t*$ can be written as functionals of the distribution of the observed data $P^*$. Abstractly, define the functional by $\psi(\cdot): {\cal M} \rightarrow R$, with true value $\psi^* = \psi(P^*)$.  For sufficiently smooth functionals like those considered in this paper, it can be shown that
\begin{equation}
\psi(P) - \psi(P^*)  = \int  G(P)(o)d(P-P^*)(o) + Rem(P,P^*) 
\label{firstorder} 
\end{equation}
where (1) $G$ is an analytic object depending on $P$ and observation value $o$ with $E_P[G(P)(O)] = \int G(P)(o) dP(o)=0$ and $Var_P[G(P)(O)] = \int G(P)(o)^2 dP(o) < \infty$ and (2)  $Rem(P,P^*)$ is a "second-order" remainder term that  involves products of differences between $P$ and $P^*$ (or their components) and tends to zero as $P$ tends to $P^*$.  The object $G$ can be viewed as a ``gradient" as it measures, at any given data-generating distribution $P$, the change in $\psi(P)$ following a slight perturbation of $P$.  Equation (\ref{firstorder}) is called a von Mises expansion, a distributional version of a Taylor expansion \citep{bickel1993efficient,carone2014higher,robins2017minimax}.
 
Provided \eqref{firstorder} holds and for a given estimator $\widehat{P}$ of $P^*$, 
\begin{equation}
\psi(\widehat{P})-\psi(P^*) =  \underbrace{- \int  G(\widehat{P})(o) dP^*(o)}_{\mbox{Bias}}+Rem(\widehat{P},P^*), 
\end{equation}
where the first term on the right hand side represents the bias of the plug-in estimator. One way to estimate the bias is by $- \int  G(\widehat{P})(o) dP_n(o)$, where $P_n$ denotes the empirical distribution based on $O_1,O_2,\ldots,O_n$, to produce the corrected plug-in or one-step estimator
\[
\widehat{\psi} = \psi(\widehat{P}) + \int  G(\widehat{P})(o) dP_n(o).
\]
This estimator ``uses the same data twice" and asymptotic theory will require the imposition of potentially restrictive Donkser conditions \citep{van2000asymptotic}.  

An alternative approach that works under weaker conditions is sample splitting \citep{chernozhukov2016double,robins2008higher,zheng2010asymptotic}.  The idea is to randomly split the observed data into $K$ (approximately) equally-sized disjoint sets.   Let $S_i$ be the split membership of the $i$th observation (i.e, $S_i \in \{ 1,\ldots, K\}$).  Let $\hat{P}^{(-k)}$ be the estimator of $P^*$ based on all the observed data except that of $k$th split, $P_{n_k}^{(k)}$ be the empirical distribution based on the $n_k$ ($\approx n/K$) observations in the $k$th split, and 
\[
\widehat{\psi}^{(k)} = \psi(\widehat{P}^{(-k)}) + \int  G(\widehat{P}^{(-k)})(o) dP^{(k)}_{n_k}(o).
\]
The sample splitting estimator of $\psi^*$ is $\widetilde{\psi} = \frac{1}{K} \sum_{k=1}^K \widehat{\psi}^{(k)}$.  To understand the behavior of this estimator, we can write $\widetilde{\psi}-\psi^*$ as
\[
\frac{1}{K} \sum_{k=1}^K \left\{ \frac{1}{n_k} \sum_{S_i=k} G(P^*)(O_i) +
\underbrace{\int  \left\{ G(\widehat{P}^{(-k)})(o) - G(P^*)(o) \right\} d( P^{(k)}_{n_k}  - P^*)(o)}_{\mbox{Term $R_{1k}$}} + \underbrace{Rem(\widehat{P}^{(-k)},P^*)}_{\mbox{Term $R_{2k}$}} \right\} \\
\]  
Using the sample splitting lemma of \cite{kennedy2018sharp}, $\sqrt{n_k} R_{1k}$ will be \newline $O_{P^*} \left( \norm{G(\widehat{P}^{(-k)}) - G(P^*) }_{L_2} \right)$, where 
$\norm{f}_{L_2}  = \sqrt{E^*[ f(O)^2 ]}$.
If  $\norm{G(\widehat{P}^{(-k)}) - G(P^*) }_{L_2}$ converges in probability to zero (Regularity Condition 1),  then $\sqrt{n_k} R_{1k}$  will be $o_{P^*}(1)$.  Further, if  $\widehat{P}^{(-k)}$ is a sufficiently well-behaved estimator of $P^*$ (i.e., converges at rates faster than $n^{1/4}$) and $R_{2k}$ is second-order (Regularity Condition 2), then $\sqrt{n_k} R_{2k}$ will be $o_{P^*}(1)$.  With $n_k = n/K$,
\[
\sqrt{n} ( \widetilde{\psi}-\psi^* ) = \frac{1}{\sqrt{n}} \sum_{i=1}^n G(P^*)(O_i) + \underbrace{\frac{1}{\sqrt{K}} \sum_{k=1}^K \sqrt{n_k} R_{k}}_{o_{P^*}(1)}
\]
where $R_k = R_{1k} + R_{2k}$.   Note that the second term on the right hand side will be $o_{P^*}(1)$ since it is  a finite sum of $o_{P^*}(1)$ terms times a constant.  The first term on the right hand side will converge in distribution, by the central limit theorem for i.i.d. data, to a normal random variable with mean zero and variance $E[ G(P^*)(O)^2 ]$.  This shows that 
$\widetilde{\psi}$ is asymptotically linear with influence function $G(P^*)(O)$.   The asymptotic variance of $\widetilde{\psi}$ can be estimated by $\frac{1}{n} \sum_{k=1}^K \sum_{i: S_i=k} G(\widehat{P}^{(-k)})(O_i)^2$.  Importantly, Equation (\ref{firstorder}) coupled with the asymptotically linearity of $\widetilde{\psi}$ will imply (under mild regularity conditions) that 
$\widetilde{\psi}$ is a "regular" \citep{newey1990semiparametric}.

Although there may be many choices of $G$ that satisfy (\ref{firstorder}), efficiency theory motivates the use of the canonical gradient, often called the efficient influence function, in the construction of the above estimator. The resulting estimator is then not only asymptotically linear but also asymptotically efficient relative to model $\mathscr{M}$. The canonical gradient can be obtained by projecting any other gradient onto the tangent space, defined at each $P\in\mathscr{M}$ as the closure of the linear span of all score functions of regular one-dimensional parametric models through $P$. A comprehensive treatment of efficiency theory can be found in \cite{pfanzagl1982lecture} and \cite{bickel1993efficient}.

The model $\mathscr{M}$ will be characterized by all distributions that satisfy the specified identification assumptions.   Depending on the estimand and assumptions, the efficient influence curve will depend on $P^*$ through a subset of $\pi^*_{t1}$ (known), $\lambda^*_1$, $\lambda_1^*(X)$, $\pi^*_t(0,X)$, $\tau_t^*(X)$, $\tau^*_t(1,X)$, $\tau^*_t(0,X)$.   The parameter $\lambda_1^*$ is estimable at $n^{1/2}$ rates by $\widehat{\lambda}_1 = \sum_i R_i/ n$. In order to estimate the functions of $X$ at fast enough rates, we will model them using generalized additive models (GAMs) in our simulations and application. We also use the rates of convergence for (GAMs) to illustrate our theoretical results. However, any desired method can be used as long as it provides the required rates of convergence.  It is important to note that these models will induce a model for 
$\pi_t^*(X)$, since 
$\pi_t^*(X) = \lambda_1^*(X) \pi_{t1}^* + (1-\lambda_1^*(X)) \pi_t^*(0,X)$.    Under appropriate smoothness conditions (i.e., the functions of the continuous variables in the generalized additive models have two derivatives), the parameters of the models for $\lambda_1^*(X)$, $\pi^*_t(0,X)$, $\tau_t^*(X)$, $\tau^*_t(1,X)$, and $\tau^*_t(0,X)$ (and thus,  
$\pi_t^*(X)$) will be estimable at $n^{2/5}$ rates \citep{horowitz2009semiparametric}.  
Where appropriate, we will comment on robustness of our estimators to model misspecification.

We will use semiparametric theory to compute the most efficient influence function in model $\mathscr{M}$ and use this to motivate the construction of asymptotically linear (AL) estimators $\mu_t^*$ and $\nu_t^*$.  Then $\Delta^*_{CC}$ and $\Delta^*_{RCT}$ can be estimated by the corresponding linear combinations of estimators for $\mu_1^*$, $\mu^*_{0}$, $\nu_{1}^*$ and $\nu^*_{0}$; linear combinations of AL estimators are AL.  The key elements of the proofs of the theorems that appear in the next two subsections can be found in the Appendix.


\subsection{Comprehensive Cohort Causal Effect} \label{cc}

 
 \begin{thm}
 Under (A1) and (A2),  $\mu_t^*$ is identified via the following formulae: 
\[
\mu_t^* = E^* \left[ \frac{ I(T=t) Y}{\pi^*_t(R,X)}\right] = E^*[ \tau_t^*(R,X) ]
\]
The optimal influence function for  $\mu_t^*$  is 
\begin{equation}
G^{(A1,A2)}_{\mu_t*}(P^*)(O)  = \frac{I(T=t) Y}{\pi^*_t(R,X)} + \left\{ 1- \frac{ I(T=t) }{\pi^*_t(R,X)} \right\} \tau_t^*(R,X) - \mu_t^*.
\end{equation}
The influence function $G^{(A1,A2)}_{\mu_t*}(P^*)(O) $ and resulting split-sampling estimator $\widetilde{\mu}^{(A1,A2)}_t$ have the following properties:
\vspace{-\topsep}
\begin{itemize}
\item $G^{(A1,A2)}_{\mu_t*}(P^*)(O)$ is robust in the sense that it is mean zero if (i) $\tau^*_t(0,X)$ is replaced by any $\tau^{\dagger}_t(0,X)$ {\em or} (ii) $\pi_t^*(0,X)$ is replaced by any $\pi_t^\dagger(0,X)$;  it has mean zero if  $\tau^*_t(1,X)$ is replaced by any $\tau^{\dagger}_t(1,X)$;
\item $\widetilde{\mu}^{(A1,A2)}_t$ will be consistent provided the models for $\tau^*_t(0,X)$ {\em or} the model for $\pi_t^*(0,X)$ is correctly specified; the model for $\tau^*_t(1,X)$ need not be correctly specified;
\item Under correct specification of models for $\tau^*_t(0,X)$ and $\pi_t^*(0,X)$ and a possibly misspecified model for $\tau^*_t(1,X)$, $\widetilde{\mu}^{(A1,A2)}_t$ will be regular and asymptotically linear with influence function $G^{(A1,A2)}_{\mu_t*}(P^{*\dagger})(O)$, where $P^{* \dagger}$ is the true distribution of the observed data with $\tau_t^*(1,X)$ replaced with $\tau^{\dagger}(1,X)$ (asymptotic limit of $\widehat{\tau}^{(-k)}_t(1,X)$).
\end{itemize}
\end{thm}
 The estimator $\widetilde{\mu}^{(A1,A2)}_t$ uses treatment, outcome and covariate data from the entire CCS. The influence function is similar to the one discussed by \cite{scharfstein1999adjusting}, \cite{bang2005doubly} and \cite{funk2011doubly}.



\begin{thm}
Under (A1) and (A3),  $\mu_t*$ is identified via the following formulae: 
\[
\mu_t^* = E^* \left[ \frac{ R I(T=t) Y}{ \lambda^*_1(X) \pi_{t1}^*}\right] = E^*[\tau_t^*(1,X)  ]
\]
The optimal influence function for  $\mu_t^*$  is 
\begin{equation}
G^{(A1,A3)}_{\mu_t*}(P^*)(O) =  \frac{ R I(T=t) Y}{ \lambda^*_1(X) \pi_{t1}^*} + \left\{ 1- \frac{ R I(T=t) }{ \lambda_1^*(X) \pi_{t1}^*}  \right\} \tau_t^*(1,X)  - \mu_t^*
\end{equation}
The influence function $G^{(A1,A3)}_{\mu_t*}(P^*)(O)$ and resulting split-sampling estimator $\widetilde{\mu}^{(A1,A3)}_t$ have the following properties:
\vspace{-\topsep}
\begin{itemize}
\item $G^{(A1,A3)}_{\mu_t*}(P^*)(O)$ is doubly robust in the sense that it is mean zero even if  (i) $\lambda_1^*(X)$ is replaced by any  $\lambda_1^{\dagger}(X)$ {\em or} (ii) $\tau_t^*(1,X)$ is replaced by any $ \tau_t^\dagger(1,X)$; 
\item $\widetilde{\mu}^{(A1,A3)}_t$ is consistent provided the model for $\lambda_1^*(X)$ or the model for $\tau_t^*(1,X)$ is correctly specified;
\item Under correct specification of models for $\lambda_1^*(X)$ and $\tau_t^*(1,X)$, $\widetilde{\mu}^{(A1,A3)}_t$ will be regular and asymptotically linear with influence function $G^{(A1,A3)}_{\mu_t*}(P^*)(O)$.
\end{itemize}
\end{thm}
The estimator $\widetilde{\mu}^{(A1,A3)}_t$ uses treatment/outcome/covariate data for all individuals in the RCT and covariate data for individuals in the OBS. The influence function is similar to the one discussed by \cite{dahabreh2018transporting}.

\begin{thm}
Under (A1), (A2) and (A3),  $\mu_t*$ is identified via the following formulae: 
\[
\mu_t^* = E^* \left[ \frac{ I(T=t) Y}{ \pi^*_t(X)}\right] = E^*[\tau_t^*(X)  ]
\]
The optimal influence function for  $\mu_t^*$  is 
\begin{equation}
G^{(A1,A2,A3)}_{\mu_t*}(P^*)(O) =  \frac{I(T=t) Y}{\pi^*_t(X)} + \left\{ 1- \frac{ I(T=t) }{\pi^*_t(X)} \right\} \tau_t^*(X) - \mu_t^*
\end{equation}
The influence function $G^{(A1,A2,A3)}_{\mu_t*}(P^*)(O)$ and resulting split-sampling estimator $\widetilde{\mu}^{(A1,A2,A3)}_t$ have the following properties:
\vspace{-\topsep}
\begin{itemize}
\item $G^{(A1,A2,A3)}_{\mu_t*}(P^*)(O)$ is doubly robust in the sense that it is mean zero even if  (i) $\pi_t^*(X)$ is replaced by any  $\pi_t^{\dagger}(X)$ {\em or} (ii) $\tau_t^*(X)$ is replaced by any $ \tau_t^\dagger(X)$; 
\item $\widetilde{\mu}^{(A1,A2,A3)}_t$ is consistent provided the model for $\pi_t^*(X)$ or the model for $\tau_t^*(X)$ is correctly specified;
\item Under correct specification of models for $\pi_t^*(X)$ and $\tau_t^*(X)$, $\widetilde{\mu}^{(A1,A2,A3)}_t$ will be regular and asymptotically linear with influence function $G^{(A1,A2,A3)}_{\mu_t*}(P^*)(O)$.
\end{itemize}
\end{thm}
The estimator $\widetilde{\mu}^{(A1,A2,A3)}_t$, uses treatment, outcome and covariate data from the entire CCS.  Under Assumptions (A1), (A2) and (A3) and correct model specification, $\widetilde{\mu}^{(A1,A2,A3)}_t$ will be as or more efficient than $\widetilde{\mu}^{(A1,A2)}_t$ and $\widetilde{\mu}^{(A1,A3)}_t$.
 
\subsection{Randomized Trial Causal Effect} \label{rct}

\begin{thm}
Under (A1), $\nu^*_t$ is identified via the following formulae: 
\[
\nu_t^* = E^* \left[ \frac{ R I(T=t) Y}{\lambda_1^*\pi_{t1}^*}\right] = E^*\left[ \frac{ R \tau_t^*(1,X)}{\lambda_1^*}  \right]
\]
The optimal influence function for  $\nu_t^*$  is
\begin{equation}
G^{(A1)}_{\nu_t*}(P^*)(O)   =  \frac{R}{\lambda_1^*}   \left\{ \frac{ I(T=t) Y}{ \pi_{t1}^*} + \left\{ 1- \frac{I(T=t) }{\pi_{t1}^*}  \right\} \tau_t^*(1,X) - \nu_t^* \right\}
\end{equation}
The influence function $G^{(A1)}_{\nu_t*}(P^*)(O)$ and resulting split-sampling estimator $\widetilde{\nu}^{(A1)}_t$ have the following properties:
\vspace{-\topsep}
\begin{itemize}
\item $G^{(A1)}_{\nu_t*}(P^*)(O)$ is robust in the sense that it has mean zero even if  $\tau_t^*(1,X)$  is replaced by $\tau_t^\dagger(1,X)$.
\item $\widetilde{\nu}^{(A1)}_t$ will be consistent even if the model for $\tau_t^*(1,X)$ is incorrectly specified. 
\item Under a possibly misspecified model for $\tau^*_t(1,X)$, $\widetilde{\nu}^{(A1)}_t$ will be regular and asymptotically linear with influence function $G^{(A1)}_{\nu_t*}(P^{*\dagger})(O)$, where $P^{* \dagger}$ is the true distribution of the observed data with $\tau_t^*(1,X)$ replaced with $\tau^{\dagger}(1,X)$ (asymptotic limit of $\widehat{\tau}^{(-k)}_t(1,X)$).
\end{itemize}
\end{thm}
The estimator $\widetilde{\nu}^{(A1)}_t$ uses treatment/outcome/covariate data for all individuals in the RCT.   It will be consistent even if the model for $\tau_t^*(1,X)$ is incorrectly specified. The influence function is the same as that discussed in Chapter 13 of \cite{tsiatis2006semiparametric}.  Imposing either (A2) or (A3), above and beyond (A1), does not result in any efficiency improvement.


\begin{thm}
Under (A1), (A2) and (A3), $\nu^*_t$ is identified via the following formulae: 
\[
\nu_t^* = E^* \left[ \frac { I(T=t) \lambda_1^*(X)  Y}{ \lambda_1^* \pi^*_t(X)}  \right] =  E^* \left[ \frac{ \lambda_1^*(X) \tau_t^*(X) }{\lambda_1^*}  \right]  
\]
The optimal influence function for  $\nu_t^*$  is
\begin{equation}
G^{(A1,A2,A3)}_{\nu_t*}(P^*)(O) =  \frac { I(T=t) \lambda_1^*(X)  Y}{ \lambda_1^* \pi_t^*(X) } 
+ \left\{ R - \frac{ I(T=t) \lambda_1^*(X)  }{\pi_t^*(X) } \right\} \frac{ \tau_t^*(X)}{\lambda_1^*} - \frac{R}{\lambda_1^*} \nu_t^*
\end{equation}
The influence function $G^{(A1,A2,A3)}_{\nu_t*}(P^*)(O)$ and resulting split-sampling estimator $\widetilde{\nu}^{(A1,A2,A3)}_t$ have the following properties:
\vspace{-\topsep}
\begin{itemize}
\item $G^{(A1,A2,A3)}_{\nu_t*}(P^*)(O)$ is robust in the sense that it has mean zero even if (i) $\tau_t^*(X)$ is replaced by $\tau^*_t(X)$ {\em} or (ii)  $\delta^*(X) = (\lambda_1^*(X),\pi_t^*(0,X))$ is replaced by $\delta_t^\dagger(X) \not = \delta^*_t(X)$. 
\item $\widetilde{\nu}^{(A1,A2,A3)}_t$ will be consistent provided the model for $\tau_t^*(X)$ is correctly specified {\em or} the models for $\lambda_1^*(X)$ {\em and} $\pi_t^*(0,X)$ are correctly specified.
\item Under correct specification of models for $\pi_t^*(X)$ and $\tau_t^*(X)$, $\widetilde{\nu}^{(A1,A2,A3)}_t$ will be regular and asymptotically linear with influence function $G^{(A1,A2,A3)}_{\nu_t*}(P^*)(O)$.
\end{itemize}
\end{thm}
The estimator $\widetilde{\nu}^{(A1,A2,A3)}_t$ 
uses treatment/outcome/covariate data from the entire CCS.  

\subsection{Remarks:} Although the estimators $\widetilde{\mu}_t^{(A1,A2)}$, $\widetilde{\mu}_t^{(A1,A3)}$ and $\widetilde{\nu}_t^{(A1,A2,A3)}$ are doubly robust with respect to consistency, achieving doubly robust inference (i.e., constructing confidence intervals with nominal coverage even if one of the modeling conditions is misspecified) is not straightforward in nonparametric settings. We refer to \cite{van2014targeted} and \cite{benkeser2017doubly} for recent progress on this problem. An alternative would be to use parametric models, for which the contribution to the variance is more straightforward to derive under misspecification. Although this would allow doubly robust inference, it would come at the cost of making much more restrictive assumptions on the nuisance functions. Since our approach only requires relatively slow second-order rate conditions (e.g., $n^{1/4+\epsilon}$ rates), the nuisance functions can be modeled flexibly, thereby reducing the risk of misspecification (making doubly robust inference a less crucial goal).

\section{Analysis of BARI} \label{chap2:sec:data}

BARI was designed to compare survival in patients receiving either PTCA or CABG. As summarized in \cite{brooks2000predictors}, a comprehensive cohort design was adopted and included 3,839 patients who had severe angina or ischemia and multivessel coronary artery disease suitable for initial revascularization by either PTCA or CABG, and who were willing to be followed up. Among these patients, 1,829 patients consented to randomization and entered a randomized trial. The remaining 2,010 patients refused randomization but agreed to participate in the BARI registry, in which patients could choose their initial treatment in consultation with their physician. The follow-up plan was similar for randomized (RCT) and registry (OBS) patients.   

In our re-analysis of the BARI study, we excluded 196 registry patients who did not receive treatment within 3 months of study entry as well as 33 RCT patients who did not receive their assigned treatment.  Our analysis utilized information on 12 baseline variables: age, sex, highest level of education, systolic blood pressure, diastolic blood pressure, qualifying symptoms (unstable angina/MI vs. other), number of diseased vessels (three vs. less than three), proximal left anterior descending disease, prior myocardial infarction, diabetes (no, with treatment, without treatment),  current smoking, hypertension. All variables are categorical except age, systolic blood pressure and diastolic blood pressure which are continuous. Patients missing at least one of these covariates ($n=149$) were excluded.  Finally, four patients with less than five years of follow-up and known to be alive at last follow-up were excluded.  Thus, our analysis included data on 3,457 patients (1,695 from OBS and 1,762 from RCT). Among RCT patients, 888 (50.4\%) received PTCA and among OBS patients, 1,108 (65.4\%) chose and received PTCA.   

Table \ref{chap2:tb:data:baseline} presents the baseline characteristics of patients, stratified by study type and treatment.  The table shows that, in aggregate, patients who enrolled in the RCT tended to be slightly sicker than those who enrolled in the OBS.  Further stratifying by treatment group reveals more stark differences between RCT and OBS patients, especially with respect to number of diseased vessels and proximal left anterior descending disease.  Since consenting to participate in an RCT is a personal patient decision, it is possible that there may be differences between RCT and OBS patients with respect to unmeasured factors such as self-efficacy and health behaviors (i.e., (A3) may be violated).  Within the RCT, the two treatments were well balanced with respect to the baseline factors.  Within the OBS, we see that CABG patients tend to have more severe disease than PTCA patients. 
Within the OBS, the selection of PTCA as compared to CABG is largely based on the patients' coronary anatomy and symptoms.    Physicians play a critical role in advising patients regarding procedure selection, and physician decisions are generally based on objective clinical and angiographic criteria. It is therefore plausible that we have accounted for the most important prognostic factors that are related to treatment selection within the OBS (i.e., (A2) is reasonable). In the RCT, the proportion of patients who died by five years was 13.40\% and 10.30\% in the PTCA and CABG arms, respectively.  In the OBS, the death rate was uniformly lower with 7.94\% and 8.86\% of PTCA and CABG patients dying, respectively.

\begin{table}
\centering
\caption{Baseline characteristics of the BARI study participants} \label{chap2:tb:data:baseline}
\begin{tabular}{lrrrrrr}
& \multicolumn{3}{c}{RCT} & \multicolumn{3}{c}{OBS}  \\ \cline{2-7}
& Total & PTCA & CBAG & Total & PTCA & CBAG \\ \cline{2-7}
$n$ & 1762 & 888 & 874 & 1695 & 1108 & 587 \\ \hline 
Age (Mean) & 60.91 & 61.20 & 60.62 & 60.97 & 60.47 & 61.90\\
Male (\%) & 73.21 & 72.52 & 73.91 & 74.61 & 74.19 & 74.11 \\
Highest Education Level \\
\; \; \; High School (\%) & 50.45 & 50.11 &50.80 & 47.55 & 46.39 & 49.74 \\
\; \; \; Some College (\%) & 18.10 & 17.91 & 18.31 & 20.24 & 20.67 & 19.42 \\
\; \; \; College/Professional (\%) & 10.56 & 11.04 & 10.07 & 21.00 & 21.12 & 20.78 \\
SBP (Mean) & 129.78 & 128.55 & 131.02 & 129.79 & 129.12 & 131.05 \\
DBP (Mean) & 75.93 & 75.27 & 76.59 & 75.68 & 75.42 & 76.19 \\
Qualifying Symptoms \\
\; \; \; Unstable Angina/MI & 68.39 & 67.00 & 69.79 & 68.38 & 67.24 & 70.53 \\
Three Diseased Vessels (\%) & 39.56 & 38.29 & 40.85 & 36.99 & 29.69 & 50.77\\
Proximal Left Anterior \\
\; \; \; Descending Disease & 40.24 & 40.65 & 39.82 & 36.58 & 31.50 & 46.17\\
Prior Myocardial Infarction & 54.26 & 53.49 & 55.03 & 50.03 & 49.91 & 50.26 \\
History of Diabetes \\
\; \; With Treatment (\%) & 19.01 & 18.58 & 19.45 & 16.81 & 15.52 & 19.25  \\
\; \; Without Treatment (\%) & 4.94 & 5.07 & 4.81 & 4.60 & 4.15 & 5.45 \\
Current Smoking (\%) & 25.43 & 26.35 & 24.49 & 20.83 & 23.01 & 16.70\\
Hypertension (\%) & 48.64 & 48.87 & 48.40 & 48.02 & 45.84 & 52.13 \\ \hline
\end{tabular}
\end{table}

In our analysis, $R$ denotes the randomization consent indicator, $T$ denotes the indicator of receiving PTCA, $Y$ denotes the indicator of death by the end of 5 years and $X$ is the vector of baseline variables. We evaluated  (A2) and (A3) by fitting, separately by treatment group, a generalized additive logistic regression model for the probability of dying by the end of 5 years as a function of the randomization consent indicator, the categorical variables (treated as factors) as well as smooth functions of age, systolic blood pressure and diastolic blood pressure.   In the PTCA model, the estimated conditional odds ratio of dying for patients enrolled in the RCT versus OBS was 1.68 (95\% CI: 1.09 to 2.58) providing statistical evidence against (A2) and (A3).  In the CABG model, the estimated conditional odds ratio of dying for patients enrolled in the RCT versus OBS was 1.13 (95\%: 0.66 to 1.96), yielding equivocal evidence against (A2) and (A3).

Despite evidence against  (A2) and (A3), we proceeded, for illustrative purposes, to estimate $\mu_t^*$ and $\nu_t^*$ under various combinations of assumptions. Towards this end, we fit generalized additive logistic regression models for $\lambda_1^*(X)$, $\pi_1^*(0,X)$, $\tau_1^*(X)$, $\tau_0^*(X)$, $\tau^*_1(1,X)$, $\tau^*_0(1,X)$, $\tau^*_1(0,X)$, and $\tau^*_0(0,X)$.  Our sample splitting estimators were based on $K=5$ splits.  Table \ref{chap2:tb:data:death} displays the estimated comprehensive cohort and randomized trial causal effects (along with standard errors and 95\% Wald-based confidence intervals) under the different assumptions.  

For the comprehensive cohort causal effect, the estimator under (A1) and (A2) has greater precision than that under (A1) and (A3). Comparatively speaking, the latter estimator is further from the null with an associated 95\% confidence interval that excludes the null.   The estimator that imposes all three assumptions was closer to the estimator under  (A1) and (A2) than under (A1) and (A3).   For the randomized trial causal effect, the estimator under (A1), (A2) and (A3) has, as expected, greater precision than that under (A1). The former estimator is further from the null, with a 95\% confidence interval that excludes the null.  Given the evidence against (A2) and (A3) in the PTCA arm, the only trustworthy estimator is the randomized trial causal effect.  Thus, our analysis cannot provide evidence that effect seen in the RCT generalizes to the entire cohort.  Our findings are largely consistent with the findings of \cite{frye1996comparison} and \cite{feit2000long}.


\begin{table}
\centering
\caption{Comprehensive cohort and randomized trial causal effect of PTCA vs. CABG on 5-year mortality (\%) for BARI} \label{chap2:tb:data:death}
\begin{tabular}{lllll} 
\multicolumn{5}{c}{Comprehensive Cohort Causal Effect} \\ \hline
Assumptions & Parameter & Estimate & S.E. & 95\% C.I. \\ \hline
 (A1), (A2) &  $\mu_1^*$ &  10.85\% & 0.72\% & 9.43\% to 12.27\%\\
 & $\mu_0^*$ & 9.16\% & 0.82\% & 7.55\% to 10.77\% \\ \cline{2-5}
 & $\Delta^*_{CC}$ & 1.68\% & 1.09\% & -0.46\% to 3.83\% \\ \hline
 (A1), (A3) &  $\mu_1^*$ &  13.26\% & 1.15\% & 11.00\% to 15.52\%\\
 & $\mu_0^*$ & 10.00\% & 1.04\% & 7.96\% to 12.03\% \\ \cline{2-5}
 & $\Delta^*_{CC}$ & 3.27\% & 1.55\% & 0.22\% to 6.31\% \\ \hline 
 (A1), (A2), (A3) &  $\mu_1^*$ &  10.65\% & 0.70\% & 9.28\% to 12.03\%\\
 & $\mu_0^*$ & 9.57\% & 0.76\% & 8.08\% to 11.06\% \\ \cline{2-5}
 & $\Delta^*_{CC}$ & 1.08\% & 1.04\% & -0.95\% to 3.11\% \\ \hline \\ \\
\multicolumn{5}{c}{Randomized Trial Causal Effect} \\ \hline
Assumptions & Parameter & Estimate & S.E. & 95\% C.I. \\ \hline
 (A1) &  $\nu_1^*$ &  13.31\% & 1.13\% & 11.08\% to 15.53\%\\
 & $\nu_0^*$ & 10.25\% & 1.02\% & 8.25\% to 12.25\% \\ \cline{2-5}
 &  $\Delta^*_{RCT}$ & 3.06\% & 1.53\% & 0.06\% to 6.05\% \\ \hline
 (A1), (A2), (A3) &  $\nu_1^*$ &  11.00\% & 0.77\% & 9.49\% to 12.51\%\\
 & $\nu_0^*$ & 9.92\% & 0.80\% & 8.36\% to 11.49\% \\ \cline{2-5}
 &  $\Delta^*_{RCT}$ & 1.07\% & 1.11\% & -1.10\% to 3.25\% \\ \hline \\ \\
\end{tabular}
\end{table}

 \section{Simulation studies} \label{chap2:sec:simu}
 
In the Appendix D, we present simulation studies, motivated by the BARI study, to evaluate the performance of the five proposed estimators.   We consider three simulation studies,  corresponding to the three sets of assumption represented in Figure \ref{fig:dag}.  For each set of assumptions, we consider various model misspecification scenarios. Since we have assumed that (A1) holds for all scenarios and $\widetilde{\nu}_t^{(A1)}$ is immune to model misspecification, all simulations demonstrate that it is unbiased, with excellence correspondence between the average standard error and standard deviation of parameter estimates and coverage of 95\% confidence intervals close to their nominal level.  The other estimators performed consistent with theory.

\section{Conclusion and Discussion} \label{chap2:sec:con}

In this paper, we presented methods for estimating causal effects of a binary treatment in the CCS design.  We discussed three estimators of the comprehensive cohort causal effect and two estimators for the randomized trial causal effect.   

For the comprehensive cohort casual effect estimators, our data analysis and simulation study suggest that the estimator that uses  (A1) and (A2) is more efficient than
 the estimator that utilizes (A1) and (A3).  In fact, it is possible to manufacture data generating scenarios where the opposite is true.  The choice between the estimators should ultimately rely on substantive tenability of the underlying assumptions.  
 
For the randomized trial causal effect estimators, $\widetilde{\nu}_t^{(A1)}$ has enviable robustness properties.  If, however, (A2) and (A3) are tenable and either a model for $\tau_t^*(X)$  {\em or} models for $\lambda_1^*(X)$ {\em and} $\pi_t^*(0,X)$ can be correctly specified, sizable efficiency gains can be achieved by using $\widetilde{\nu}_t^{(A1, A2,A3)}$.  In settings where it can be difficult to enroll the requisite number of patients for an adequately powered randomized trial and one is interested in the randomized trials causal effect, the estimator may be a viable way of obtaining more precise inferences.  More detailed study of the bias-variance tradeoff associated with using data from OBS to estimate the RCT effect is warranted.

Finally, for both estimands, the methods in this paper need to be extended to deal with non-compliance, missing data, time-varying treatments and censored outcomes.



%




\newpage

\bibliographystyle{unsrtnat} 
\bibliography{reference_CCS.bib}

\newpage



\noindent {\bf \Large Appendices}

\vspace*{0.1in}

\begin{appendices}

In what follows, we define $\pi^*_t(0,X,Y_1,Y_0)  = P[T=t | R=0, X, Y_1, Y_0]$. It is also useful to note that any observed data random variable can be written as:
$RT h_1(X,Y_1) + R(1-T) h_2(X,Y_0) + (1-R) T h_3(X,Y_1) + (1-R)(1-T) h_4(X,Y_0)$ for some functions $h_1(X,Y_1)$, $h_2(X,Y_0)$,  $h_3(X,Y_1)$,  $h_4(X,Y_0)$.

\section{Orthogonal Complement of Tangent Spaces}

\subsection{Assumption (A1)}

The observed data tangent space is:
\[
{\mathscr{T}}  =  A \oplus B
\]
where
\[
A = \{ E[ a(R,X,Y_1,Y_0) | O] : E[a(R,X,Y_1,Y_0)]=0 \}
\]
\[
B = \{ E[ (1-R) (T - \pi^*_1(0,X,Y_1,Y_0) ) b(X, Y_1, Y_0) | O] : b(X, Y_1, Y_0) \}
\]
The orthogonal complement of ${\mathscr{T}}$ is ${\mathscr{T}}^{\perp} = A^{\perp} \cap B^{\perp}$.  Notice that
\begin{eqnarray*}
A^{\perp} & = & \{ R T h_1(X,Y_1) + R(1-T) h_2(X,Y_0) : \pi^*_{11} h_1(X,Y_1) + \pi^*_{01} h_2(X,Y_0)=0\} \oplus \\
&& \left\{ (1-R) T h_3(X,Y_1) + (1-R) (1-T) h_4(X,Y_0) : \right. \\
&& \hspace*{0.5in} \left. \pi_1^*(0,X,Y_1,Y_0) h_3(X,Y_1) + \pi_0^*(0,X,Y_1,Y_0) h_4(X,Y_0) = 0 \right\} \\
& = & \{ R ( T - \pi^*_{11} ) h (X):  h(X) \} \oplus \\
&& \left\{ (1-R) T h_3(X,Y_1) + (1-R) (1-T) h_4(X,Y_0) : \right. \\
&& \hspace*{0.5in} \left. \pi_1^*(0,X,Y_1,Y_0) h_3(X,Y_1) + \pi_0^*(0,X,Y_1,Y_0) h_4(X,Y_0) = 0 \right\}
\end{eqnarray*}
We must now find elements of $A^{\perp}$ that are in $B^{\perp}$. We first note that $R ( T - \pi^*_{11} ) h (X)$ is in $B^{\perp}$ for all $h(X)$.  We now seek additional conditions on $h_3(X,Y_1)$ and $h_4(X,Y_1)$ such that $(1-R) T h_3(X,Y_1) + (1-R) (1-T) h_4(X,Y_0)$ is in  $B^{\perp}$.  It must be the case that 
\begin{eqnarray*}
0 & = &  E[ (T - \pi^*_1(0,X,Y_1,Y_0) \{ T h_3(X,Y_1) + (1-T) h_4(X,Y_0) \} | R=0, X, Y_1,Y_0]   \\
& = &  \pi^*_1(0,X,Y_1,Y_0) \pi^*_0(0,X,Y_1,Y_0) \{ h_3(X,Y_1) -  h_4(X,Y_0) \} 
\end{eqnarray*}
This implies that $ h_3(X,Y_1) =  h_4(X,Y_0)$.  Imposing the condition on $h_3(X,Y_1)$ and $h_4(X,Y_1)$ from $A^{\perp}$, we then have that $ h_3(X,Y_1) =  h_3(X,Y_0)=0$.  Thus, ${\mathscr{T}}^{\perp} = \{ R ( T - \pi^*_{11} ) h (X):  h(X) \}$.
 
\subsection{Assumption (A1,A2)}

The observed data tangent space is:
\[
{\mathscr{T}}  =  A'  \oplus B'
\]
where
\[
A' =  \{ E[ a(R,X,Y_1,Y_0) | O] : E[a(R,X,Y_1,Y_0)]=0 \} 
\]
\[
B' = \{ (1-R) (T - \pi^*_1(0,X) ) b(X) : b(X) \}
\]
Here, ${\mathscr{T}}^{\perp} = A^{'\perp} \cap B^{'\perp}$.  Notice that
\begin{eqnarray*}
A^{'\perp} & = & \{ R T h_1(X,Y_1) + R(1-T) h_2(X,Y_0) : \pi^*_{11} h_1(X,Y_1) + \pi^*_{01} h_2(X,Y_0)=0\} \oplus \\
&& \left\{ (1-R) T h_3(X,Y_1) + (1-R) (1-T) h_4(X,Y_0) : \right. \\
&& \hspace*{0.5in} \left. \pi_1^*(0,X) h_3(X,Y_1) + \pi_0^*(0,X) h_4(X,Y_0) = 0 \right\} \\
& = & \{ R ( T - \pi^*_{11} ) h (X):  h(X) \} \oplus \{ (1-R) (T- \pi_1^*(0,X)) h(X) : h(X) \} 
\end{eqnarray*}
We must now find elements of $A^{'\perp}$ that are in $B^{'\perp}$. We first note that $R ( T - \pi^*_{11} ) h (X)$ is in $B^{'\perp}$ for all $h(X)$.  We now seek additional conditions on $b(X)$  such that $(1-R) (T- \pi_1^*(0,X)) h(X)$ is in  $B^{\perp}$.  Since  $(1-R) (T- \pi_1^*(0,X)) h(X) \in B$ for all $h(X)$, it must be the case that $h(X)=0$.   Thus, ${\mathscr{T}}^{\perp} = \{ R ( T - \pi^*_{11} ) h(X):  h(X) \}$.

\subsection{Assumption (A1,A3)}

The observed data tangent space is:
\[
{\mathscr{T}}  =  A^{\dagger} \oplus B^{\dagger} \oplus C^{\dagger}
\]
where
\[
A^{\dagger}= \{ E[ a(X,Y_1,Y_0) | O] : E[a(X,Y_1,Y_0)]=0 \}
\]
\[
B^{\dagger} =  \{ E[ (1-R) (T - \pi^*_1(0,X,Y_1,Y_0) ) b(X, Y_1, Y_0) | O] : b(X, Y_1, Y_0) \}
\]
\[
C^{\dagger} =  \{  (R - \lambda_1^*(X) \} c(X) : c(X) \}
\]
Here, ${\mathscr{T}}^{\perp} = A^{\dagger \perp} \cap B^{ \dagger \perp} \cap C^{ \dagger \perp} $.  Notice that
\begin{eqnarray*}
A^{\dagger \perp} & = & \left\{ R T h_1(X,Y_1) + R(1-T) h_2(X,Y_0) + \right. \\
&&  \hspace*{0.1in} (1-R) T h_3(X,Y_1) + (1-R)(1-T) h_4(X,Y_0)  : \\
&& \hspace*{0.5in} \lambda_1^*(X) \pi_{11}^* h_1(X,Y_1) +\lambda_1^*(X) \pi_{01}^* h_2(X,Y_0) +  \\
&& \hspace*{0.5in} \left. \lambda_0^*(X) \pi_1^*(0,X,Y_1,Y_0) h_3(X,Y_1) +\lambda_0^*(X) \pi_0^*(0,X,Y_1,Y_0) h_4(X,Y_0) = 0 \right\}
\end{eqnarray*}
Let's find elements of $A^{\dagger \perp}$ that are in $B^{\dagger \perp}$.  The additional condition on $h_3(X,Y_1)$ and $h_4(X,Y_0)$ is 
\begin{eqnarray*}
0 & = &  E[ (T - \pi^*_1(0,X,Y_1,Y_0) \{ T h_3(X,Y_1) + (1-T) h_4(X,Y_0) \} | R=0, X, Y_1,Y_0]   \\
& = &  \pi^*_1(0,X,Y_1,Y_0) \pi^*_0(0,X,Y_1,Y_0) \{ h_3(X,Y_1) -  h_4(X,Y_0) \} 
\end{eqnarray*}
This implies that $ h_3(X,Y_1) =  h_4(X,Y_0) = h(X)$ and
\[
\lambda_1^*(X) \pi_{11}^* h_1(X,Y_1) +\lambda_1^*(X) \pi_{01}^* h_2(X,Y_0) +  \lambda_0^*(X) h(X)  = 0
\]
This can only hold if $h_1(X,Y_1)=h_1(X)$ and $h_2(X,Y_0)=h_2(X)$. Thus,
\begin{eqnarray*}
A^{\dag \perp} \cap B^{\dag \perp}& = & \left\{ R T h_1(X) + R(1-T) h_2(X) + (1-R) h(X) \right.   : \\
&& \hspace*{0.5in} \left. \lambda_1^*(X) \pi_{11}^* h_1(X) +\lambda_1^*(X) \pi_{01}^* h_2(X) + \lambda_0^*(X) h(X) = 0 \right\}
\end{eqnarray*}
We must now find elements of $A^{\dagger \perp} \cap B^{\dagger \perp}$ that are in $C^{\dagger \perp}$. The additional condition on $h_1(X)$, $h_2(X)$ and $h(X)$ is
\begin{eqnarray*}
0 & = &  E[ (R - \lambda^*_1(X) \{  R T h_1(X) + R(1-T) h_2(X) + (1-R) h(X) \} | X]   \\
& = &  \lambda_1^*(X) \lambda_0^*(X)) \{ \pi^*_{11} h_1(X) +  \pi^*_{01} h_2(X) - h(X) \}
\end{eqnarray*}
This implies that $\pi^*_{11} h_1(X) +  \pi^*_{01} h_2(X) - h(X)=0$.  Further imposing the condition from $A^{\dagger \perp} \cap B^{\dagger \perp}$, we have  $h(X)=0$ and $h_2(X) = - \frac{ \pi^*_{11} }{ \pi^*_{01} } a_1(X)$.  Thus, ${\mathscr{T}}^{\perp} = \{ R ( T - \pi^*_{11} ) a (X):  a(X) \}$.

\subsection{Assumption (A1,A2,A3)}

The observed data tangent space is:
\[
{\mathscr{T}}  =  A^{\ddag} \oplus B^{\ddag} \oplus C^{\ddag}
\]
where
\[
A^{\ddag}= \{ E[ a(X,Y_1,Y_0) | O] : E[a(X,Y_1,Y_0)]=0 \}
\]
\[
B^{\ddag} = \{ (1-R) (T - \pi^*_1(0,X) ) b(X) : b(X) \}
\]
\[
C^{\ddag} =  \{  (R - \lambda_1^*(X) \} c(X) : c(X) \}
\]

Here, ${\mathscr{T}}^{\perp} = A^{\ddag \perp} \cap B^{ \ddag \perp} \cap C^{ \ddag \perp} $.  Notice that
\begin{eqnarray*}
A^{\ddag \perp} & = & \left\{ R T h_1(X,Y_1) + R(1-T) h_2(X,Y_0) + \right. \\
&&  \hspace*{0.1in} (1-R) T h_3(X,Y_1) + (1-R)(1-T) h_4(X,Y_0)  : \\
&& \hspace*{0.5in} \lambda_1^*(X) \pi_{11}^* h_1(X,Y_1) +\lambda_1^*(X) \pi_{01}^* h_2(X,Y_0) +  \\
&& \hspace*{0.5in} \left. \lambda_0^*(X) \pi_1^*(0,X) h_3(X,Y_1) +\lambda_0^*(X) \pi_0^*(0,X) h_4(X,Y_0) = 0 \right\}. \\
& = & \left\{ \left\{ \frac{RT}{\lambda_1^*(X) \pi_{11}^*} -  \frac{(1-R) T}{\lambda_0^*(X) \pi_{1}^*(0,X)}  \right\} h_1(X,Y_1) +  \right. \\
&& \left\{ \frac{R(1-T)}{\lambda_1^*(X) \pi_{01}^*} -  \frac{(1-R) (1-T)}{\lambda_0^*(X) \pi_{0}^*(0,X)}  \right\} h_2(X,Y_0) + \\
&& \left. \frac{(1-R)}{\lambda_0^*(X)}  \left\{ T - \pi_{1}^*(0,X) \right\} h(X):   h_1(X,Y_1), h_2(X,Y_0), h(X)  \right\}. 
\end{eqnarray*}
Let's find elements of $A^{\ddag \perp}$ that are in $B^{\ddag \perp}$.  The additional condition on $h_1(X,Y_1)$, $h_2(X,Y_0)$ and $h(X)$ is 
\begin{eqnarray*}
0 & = &  E \left[ (1-R) (T - \pi^*_1(0,X) )\left\{ -  \frac{T}{\lambda_0^*(X) \pi_{1}^*(0,X)}  h_1(X,Y_1) - \frac{(1-T)}{\lambda_0^*(X) \pi_{0}^*(0,X)}  h_2(X,Y_0)  +  \right. \right. \\
&&  \hspace*{1.5in}  \left. \left. \left\{ \frac{T - \pi_{1}^*(0,X)}{\lambda^*_0(X)}  \right\} h(X)  \right\} \; \vline \;  X \right] \\
& = &  h(X)  \pi^*_1(0,X) \pi^*_0(0,X)-  \pi^*_0(0,X) E[h_1(X,Y_1)|X] + \pi^*_1(0,X) E[h_2(X,Y_1)|X] 
\end{eqnarray*}
which implies that
\[
h(X) =  \frac{E[h_1(X,Y_1)|X]}{\pi^*_1(0,X) } - \frac{E[h_2(X,Y_1)|X]}{ \pi^*_0(0,X) } 
\]
Substituting this expression for $h(X)$ into $A^{\ddag \perp}$, we can now find elements in $A^{\dagger \perp} \cap B^{\dagger \perp}$ that are in $C^{\ddag \perp}$. The additional condition on $h_1(X,Y_1)$ and $h_2(X,Y_0)$ is 
\begin{eqnarray*}
0 & = &  E \left[ (R-\lambda_1^*(X)) \left\{ \left\{ \frac{RT}{\lambda_1^*(X) \pi_{11}^*} -  \frac{(1-R) T}{\lambda_0^*(X) \pi_{1}^*(0,X)}  \right\} h_1(X,Y_1)  +  \right. \right. \\
&&  \hspace*{0.5in}  \left\{ \frac{R(1-T)}{\lambda_1^*(X) \pi_{01}^*} -  \frac{(1-R) (1-T)}{\lambda_0^*(X) \pi_{0}^*(0,X)}  \right\} h_2(X,Y_0) + \\
&&  \hspace*{0.5in}  \left. \left. \left\{ \frac{1-R}{\lambda_0^*(X)} \right\} \left\{ \frac{T - \pi_{1}^*(0,X)}{\lambda^*_0(X)}  \right\}  \left\{ \frac{E[h_1(X,Y_1)|X]}{\pi^*_1(0,X) } - \frac{E[h_2(X,Y_1)|X]}{ \pi^*_0(0,X) }  \right\} \right\} \; \vline \;  X \right] \\
& = &   E[h_1(X,Y_1)|X] + E[h_2(X,Y_1)|X] 
\end{eqnarray*}
Let $\tilde{h}_1(X,Y_1) = h_1(X,Y_1)-E[h_1(X,Y_1)|X]$ and  $\tilde{h}_2(X,Y_0) = h_2(X,Y_0)-E[h_1(X,Y_0)|X]$.  After algebra, we can then write 
\begin{eqnarray*}
{\mathscr{T}}^{\perp} & = &\left\{  \left\{ \frac{RT}{\lambda_1^*(X) \pi_{11}^*} -  \frac{(1-R) T}{\lambda_0^*(X) \pi_{1}^*(0,X)}  \right\} \tilde{h}_1(X,Y_1) : E[\tilde{h}_1(X,Y_1)|X]=0 \right\}  \oplus \\
&& \left\{ \left\{ \frac{R(1-T)}{\lambda_1^*(X) \pi_{01}^*} -  \frac{(1-R) (1-T)}{\lambda_0^*(X) \pi_{0}^*(0,X)}  \right\} \tilde{h}_2(X,Y_0) : E[\tilde{h}_2(X,Y_0)|X]=0 \right\} \oplus  \\
&& \left\{ \left\{ \frac{R}{\lambda_1^*(X)} \right\}  \left\{ \frac{T}{\pi_{11}^*} - \frac{1-T}{\pi_{01}^*} \right\} h(X):  h(X) \right\}  \\
\end{eqnarray*}

\section{Efficient Influence Functions and Estimators}

For a given set of assumptions, the efficient influence for a target parameter is a naive influence function minus its projection onto the orthogonal complement of the tangent space.

\subsection{Assumption (A1)}

Under (A1), $\mu_t^*$ is not identified and so there are no influence functions to discuss.  However, $\nu_t^*$ is identified and a {\bf naive influence function} is 
\[
\phi_{\nu_t^*}(O) = \frac{R I(T=t)}{\lambda_1^* \pi_{t1}^*} (Y_t - \nu^*_t)
\]
The projection of $\phi_{\nu_t^*}(O) $ onto ${\mathscr{T}}^{\perp}$ is
\[
R ( T - \pi^*_{11} ) \frac{E[( T - \pi^*_{11} ) \phi_{\nu_t^*}(O)  | R=1,X]  }{ E[ ( T - \pi^*_{11} )^2 |R=1,X]}
\]
The projection reduces to 
\[
(-1)^{t+1} \frac{R ( T - \pi^*_{11} )}{ \lambda_1^* \pi_{t1}^*} \left\{ E[Y_t|R=1,X] - \nu_t^* \right\}
\]
So, the efficient influence function for $\nu_t^*$ is
\[
\frac{R}{\lambda_1^*} \left\{ \frac{ I(T=t)}{ \pi_{t1}^*} Y_t  + \left\{ 1- \frac{I(T=t)}{ \pi_{t1}^*}  \right\} E[Y_t|R=1,X] - \nu_t^* \right\} 
\]
Let 
\[
\widetilde{\nu}_t^{(-k)} = \frac{1}{n-n_k} \sum_{S_i \not = k} \frac{R_i \widehat{\tau}_t^{(-k)}(1,X_i) }{\widehat{\lambda}_1^{(-k)}(X_i)} 
\]
be the plug-in estimator based on data excluding the $k$th split.  The resulting estimator based on the $k$th split is
\[
\widetilde{\nu}_t^{(-k)} + \frac{1}{n_k} \sum_{S_i=k} \frac{R_i}{\widehat{\lambda}_1^{(-k)}} \left\{ \frac{ I(T_i=t)}{ \pi_{t1}^*} Y_i  + \left\{ 1- \frac{I(T_i=t)}{ \pi_{t1}^*}  \right\}\widehat{\tau}_t^{(-k)}(1,X_i) -\widetilde{\nu}_t^{(-k)} \right\} 
\]

\subsection{Assumption (A1,A2)}

Under (A1,A2), $\mu_t^*$ is identified and a naive influence function is 
\[
\phi_{\mu_t^*}(O) = \frac{I(T=t)}{\pi_t^*(R,X)} (Y_t - \mu^*_t) + (-1)^t (1-R) \frac{T - \pi^*_1(0,X)}{ \pi^*_t(0,X)} \left\{ E[Y_t|R=0,X] - \mu^*_t \right\}
\]
The projection of $\phi_{\mu_t^*}(O) $ onto ${\mathscr{T}}^{\perp}$ is
\[
R ( T - \pi^*_{11} ) \frac{E[( T - \pi^*_{11} ) \phi_{\mu_t^*}(O)  | R=1,X]  }{ E[ ( T - \pi^*_{11} )^2 |R=1,X]}
\]
The projection reduces to 
\[
(-1)^{t+1} \frac{R ( T - \pi^*_{11} )}{  \pi_{t1}^*} \left\{ E[Y_t|R=1,X] - \mu_t^* \right\}
\]
So, the efficient influence function for $\mu_t^*$ is
\[
 \frac{ I(T=t)}{ \pi_{t}^*(R,X)} Y_t  + \left\{ 1- \frac{I(T=t)}{ \pi_{t}^*(R,X)}  \right\} E[Y_t|R,X] - \mu_t^* 
 \]
The resulting estimator based on the $k$th split is
\[
\frac{1}{n_k} \sum_{S_i=k}  \left\{  \frac{ I(T_i=t)} { \widehat{\pi}^{(-k)}_{t}(R_i,X_i)} Y_i  + \left\{ 1- \frac{I(T_i=t)}{  \widehat{\pi}^{(-k)}_{t}(R_i,X_i) }  \right\} \widehat{\tau}^{(-k)}_t(R_i,X_i)  \right\},
\]
where $\widehat{\pi}^{(-k)}_{t}(R_i,X_i) = R_i \pi_{t1}^* + (1-R_i) \widehat{\pi}^{(-k)}_{t}(0,X_i)$

For $\nu_t^*$, the efficient influence curve under (A1,A2) is the same as under (A1), since ${\mathscr{T}}^{\perp}$ is the same.
 
 \subsection{Assumption (A1,A3)}

Under (A1,A3), $\mu_t^*$ is identified and a naive influence function is 
\[
\phi_{\mu_t^*}(O) = \frac{RI(T=t)}{\lambda_1^*(X)\pi_{t1}^*} \left\{ Y_t - \mu^*_t \right\}  + \left\{ 1 - \frac{R}{\lambda_1^*(X)} \right\}  \left\{ E[Y_t|R=1,X] - \mu^*_t \right\}
\]
The projection of $\phi_{\mu_t^*}(O) $ onto ${\mathscr{T}}^{\perp}$ is
\[
R ( T - \pi^*_{11} ) \frac{E[( T - \pi^*_{11} ) \phi_{\mu_t^*}(O)  | R=1,X]  }{ E[ ( T - \pi^*_{11} )^2 |R=1,X]}
\]
The projection reduces to 
\[
(-1)^{t+1} \frac{R ( T - \pi^*_{11} )}{  \pi_{t1}^* \lambda_1^*(X)} \left\{ E[Y_t|R=1,X] - \mu_t^* \right\}
\]
So, the efficient influence function for $\mu_t^*$ is
\[
\frac{RI(T=t)}{\lambda_1^*(X)\pi_{t1}^*}  Y_t  + \left\{ 1  - \frac{R I(T=t) }{\lambda_1^*(X) \pi_{t1}^*}
 \right\}  E[Y_t|R=1,X] -  \mu^*_t  
\]
The resulting estimator based on the $k$th split is
\[
\frac{1}{n_k} \sum_{S_i=k}  \left\{ \frac{R_i I(T_i=t)}{\widehat{\lambda}^{(-k)}_1(X_i)\pi_{t1}^*}  Y_i  + \left\{ 1  - \frac{R_i I(T_i=t) }{\widehat{\lambda}^{(-k)}_1(X_i) \pi_{t1}^*}
 \right\}  \widehat{\tau}^{(-k)}_t(1,X_i) \right\}
\]

For $\nu_t^*$, the efficient influence curve under (A1,A3) is the same as under (A1), since ${\mathscr{T}}^{\perp}$ is the same.

 \subsection{Assumption (A1,A2,A3)}

The projection of any observed data random variable $h(O)$ onto ${\mathscr{T}}^{\perp}$ is
\begin{align*}
 & \left\{ \frac{RT}{\lambda_1^*(X) \pi_{11}^*} -  \frac{(1-R) T}{\lambda_0^*(X) \pi_{1}^*(0,X)}  \right\} \left\{ \frac{ E \left[   \left\{ \frac{RT}{\lambda_1^*(X) \pi_{11}^*} -  \frac{(1-R) T}{\lambda_0^*(X) \pi_{1}^*(0,X)}  \right\}  h(O) \; \vline \; X, Y_1 \right]}{E \left[  \left\{ \frac{RT}{\lambda_1^*(X) \pi_{11}^*} -  \frac{(1-R) T}{\lambda_0^*(X) \pi_{1}^*(0,X)}  \right\} ^2  \; \vline \; X, Y_1 \right]} - \right.  \\
 & \left. \hspace*{2in} E \left[  \frac{ E \left[   \left\{ \frac{RT}{\lambda_1^*(X) \pi_{11}^*} -  \frac{(1-R) T}{\lambda_0^*(X) \pi_{1}^*(0,X)}  \right\}  h(O) \; \vline \; X, Y_1 \right]}{E \left[  \left\{ \frac{RT}{\lambda_1^*(X) \pi_{11}^*} -  \frac{(1-R) T}{\lambda_0^*(X) \pi_{1}^*(0,X)}  \right\} ^2  \; \vline \; X, Y_1 \right]}    \; \vline \; X \right] \right\}   +  \\
&\left\{ \frac{R(1-T)}{\lambda_1^*(X) \pi_{01}^*} -  \frac{(1-R) (1-T)}{\lambda_0^*(X) \pi_{0}^*(0,X)}  \right\} \left\{ \frac{ E \left[   \left\{ \frac{R(1-T)}{\lambda_1^*(X) \pi_{01}^*} -  \frac{(1-R) (1-T)}{\lambda_0^*(X) \pi_{0}^*(0,X)}  \right\}  h(O) \; \vline \; X, Y_0 \right]}{E \left[  \left\{ \frac{R(1-T)}{\lambda_1^*(X) \pi_{01}^*} -  \frac{(1-R) (1-T)}{\lambda_0^*(X) \pi_{0}^*(0,X)}  \right\} ^2  \; \vline \; X, Y_0 \right]} - \right.  \\
 & \left. \hspace*{2in} E \left[  \frac{ E \left[   \left\{ \frac{R(1-T)}{\lambda_1^*(X) \pi_{01}^*} -  \frac{(1-R) (1-T)}{\lambda_0^*(X) \pi_{0}^*(0,X)}  \right\}  h(O) \; \vline \; X, Y_0 \right]}{E \left[  \left\{ \frac{R(1-T)}{\lambda_1^*(X) \pi_{01}^*} -  \frac{(1-R) (1-T)}{\lambda_0^*(X) \pi_{0}^*(0,X)}  \right\} ^2  \; \vline \; X, Y_0 \right]}  \; \vline \; X \right] \right\}   +  \\
&  \left\{ \frac{R}{\lambda_1^*(X)} \right\}  \left\{ \frac{T}{\pi_{11}^*} - \frac{1-T}{\pi_{01}^*} \right\}\frac{E \left[  \left\{ \frac{R}{\lambda_1^*(X)} \right\}  \left\{ \frac{T}{\pi_{11}^*} - \frac{1-T}{\pi_{01}^*} \right\}h(O) \; \vline \; X \right] }{E \left[  \left\{ \frac{R}{\lambda_1^*(X)} \right\}^2  \left\{ \frac{T}{\pi_{11}^*} - \frac{1-T}{\pi_{01}^*} \right\}^2  \; \vline \; X \right]}
\end{align*}

Under (A1,A2, A3), a naive influence function for $\mu_t^*$ is 
\[
\phi_{\mu_t^*}(O) = \frac{RI(T=t)}{\lambda_1^*(X)\pi_{t1}^*} \left\{ Y_t - \mu^*_t \right\}  + \left\{ 1 - \frac{R}{\lambda_1^*(X)} \right\}  \left\{ E[Y_t|X] - \mu^*_t \right\}
\]
The projection of $\phi_{\mu_t^*}(O)$ onto ${\mathscr{T}}^{\perp}$ is 
\begin{align*}
&   I(T=t)  \left\{ \frac{R }{\lambda_1^*(X) \pi_{t1}^*} -  \frac{1}{\pi_{t}^*(X)}  \right\} \left\{ Y_t - E[Y_t|X] \right\} + \\
& (-1)^{t+1}   \left\{ \frac{R}{\lambda_1^*(X) \pi_{t1}^*} \right\}  \left\{ T - \pi_{11}^* \right\}  \left\{ E[Y_t|X] - \mu_t^* \right\} 
\end{align*}
So, the efficient influence function for $\mu_t^*$ is
 \[
 \frac{ I(T=t)}{ \pi_{t}^*(X)} Y_t  + \left\{ 1  - \frac{I(T=t) }{\pi_{t}^*(X)}
 \right\} E[Y_t|X] - \mu_t^*
\]
The resulting estimator based on the $k$th split is
\[
\frac{1}{n_k} \sum_{S_i=k}  \left\{   \frac{ I(T_i=t)}{ \pi^{(-k)}_{t}(X_i)} Y_i  + \left\{ 1  - \frac{I(T_i=t) }{ \pi^{(-k)}_{t}(X_i)}
 \right\} \widehat{\tau}^{(-k)}_t(X_i)   \right\},
\]
where $ \pi^{(-k)}_{t}(X_i) = \widehat{\lambda}_1^{(-k)}(X_i)  \pi_{t1}^* + \left(1-\widehat{\lambda}_1^{(-k)}(X_i) \right)  \widehat{\pi}^{(-k)}_{t}(0,X_i)$

Under (A1,A2, A3), a naive influence function for $\nu_t^*$ is 
\[
\phi_{\nu_t^*}(O) = \frac{R I(T=t)}{\lambda_1^* \pi_{t1}^*} (Y_t - \nu^*_t)
\]
The projection of $\phi_{\mu_t^*}(O) $ onto ${\mathscr{T}}^{\perp}$ is
\begin{align*}
&  I(T=t)  \left\{  \frac{R \pi_{t}^*(X)  }{\lambda_1^*(X) \pi_{t1}^*} -  1  \right\} \left\{ \frac{\lambda_1^*(X)}{ \pi_t^*(X) \lambda_1^*} \right\}  \left\{ Y_t - E[Y_t|X] \right\} + \\
& (-1)^{t+1}   \left\{ \frac{R}{\lambda_1^* \pi_{t1}^*} \right\}  \left\{ T - \pi_{11}^* \right\}  \left\{ E[Y_t|X] - \nu_t^* \right\} 
\end{align*}
So, the efficient influence function for $\nu_t^*$ is
\[
 \frac{I(T=t) \lambda_1^*(X)}{ \pi_t^*(X) \lambda_1^*} Y_t  +  \left\{ R - \frac{I(T=t) \lambda_1^*(X)}{\pi_t^*(X) } \right\} \frac{E[Y_t|X]}{\lambda_1^*}  -  \frac{R}{\lambda_1^*} \nu_t^*
 \]

Let 
\[
\check{\nu}_t^{(-k)} = \frac{1}{n-n_k} \sum_{S_i \not = k} \frac{\widehat{\lambda}_1^{(-k)}(X_i) \widehat{\tau}_t^{(-k)}(X_i) }{\widehat{\lambda}_1^{(-k)}} 
\]
be the plug-in estimator based on data excluding the $k$th split.  The resulting estimator based on the $k$th split is
\[
\check{\nu}_t^{(-k)} + \frac{1}{n_k} \sum_{S_i=k}  \left\{ \frac{I(T_i=t) \widehat{\lambda}_1^{(-k)}(X_i)}{ \pi^{(-k)}_{t}(X_i)  \widehat{\lambda}_1^{(-k)}} Y_i  +  \left\{ R_i - \frac{I(T_i=t) \widehat{\lambda}_1^{(-k)}(X_i)}{ \pi^{(-k)}_{t}(X_i)  } \right\} \frac{\widehat{\tau}^{(-k)}_t(X_i) }{\widehat{\lambda}_1^{(-k)}}  -  \frac{R_i}{\widehat{\lambda}_1^{(-k)}} \check{\nu}_t^{(-k)} \right\}
\]
where $ \pi^{(-k)}_{t}(X_i) = \widehat{\lambda}_1^{(-k)}(X_i)  \pi_{t1}^* + \left(1-\widehat{\lambda}_1^{(-k)}(X_i) \right)  \widehat{\pi}^{(-k)}_{t}(0,X_i)$

\section{Asymptotics}

\subsection{Theorem 1}

It can be shown that
\begin{eqnarray*}
Rem(P,P^*) 
& = & E^* \left[ (1-R) \left( \frac{ \pi_t^*(0,X) }{ \pi_t(0,X)} - 1 \right)  \left( \tau_t^*(0,X) -  \tau_t(0,X) \right)  \right]
\end{eqnarray*}
Letting $P$ be the estimator $\widehat{P}^{(-k)}$ and assuming  
$\widehat{\pi}^{(-k)}_t(R,X)$ is bounded away from zero with probability 1, it follows that
\begin{equation}
\label{cond2a1a2}
| Rem(\widehat{P}^{(-k)},P^*)  | \leq C^{(-k)} \underbrace{ \norm{  \widehat{\pi}^{(-k)}_t(0,X)  - \pi^*_t(0,X) } _{L_2} }_{O_{P^*}(n_k^{-2/5})} \underbrace{ \norm{ \widehat{\tau}^{(-k)}_t(0,X) -  \tau^*_t(0,X)  }_{L_2}}_{O_{P^*}(n_k^{-2/5})}
\end{equation}
where $C^{(-k)}$ is $O_{P^*}(1)$.
This implies that that Regularity Condition 2 holds and $\sqrt{n_k} R_{2k}$ is $o_{P^*}(1)$.

In terms of $R_{1k}$ is useful to notice that it is equal to $\int  \left\{ \phi(\widehat{P}^{(-k)})(o) - \phi(P^*)(o) \right\} d( P^{(k)}_{n_k}  - P^*)(o)$, where $\phi(P^*)(o) = G_{\mu_t^*}^{(A1,A2)}(P^*)(o) + \mu_t^*$. Thus, $\sqrt{n_k} R_{1k}$ will be  \newline $O_{P^*} \left( \norm{\phi(\widehat{P}^{(-k)}) - \phi(P^*) }_{L_2} \right)$.  Using the triangle and Cauchy-Schwarz inequalities, it can be shown that 
\begin{eqnarray}
\norm{ \phi(\widehat{P}^{(-k)}) - \phi(P^*) }_{L2} & \leq &  D^{(-k)}_{1} \underbrace{\norm{\widehat{\pi}^{(-k)}_t(0,X)  - \pi^*_t(0,X)  }_{L_2}}_{o_P^*(1)} +  \\
&& D^{(-k)}_{2} \underbrace{\norm{ \widehat{\tau}^{(-k)}_t(0,X) -  \tau^*_t(0,X)  }_{L_2}}_{o_P^*(1)} + \nonumber \\
&& D^{(-k)}_{3} \underbrace{\norm{ \widehat{\tau}^{(-k)}_t(1,X) -  \tau^*(1,X)  }_{L_2}}_{o_P^*(1)}. \label{cond1a1a2}
\end{eqnarray}
where $D^{(-k)}_{1} $, $D^{(-k)}_{2} $ and $D^{(-k)}_{3}$ are $O_{P^*}(1)$. Thus, $\norm{ \phi(\widehat{P}^{(-k)}) - \phi(P^*) }_{L2}$ is $o_{P^*}(1)$  (i.e., Regularity Condition 1 holds)  and $\sqrt{n_k} R_{1k}$ is $o_{P^*}(1)$.  

\subsection{Theorem 2}

It can be shown that
\begin{eqnarray*}
Rem(P,P^*) 
& = &   E^* \left[  \left( \frac{    \lambda^*_1(X) }{  \lambda_1(X) } -1 \right)  \left( \tau_t^*(1,X) -  \tau_t(1,X) \right)  \right] 
\end{eqnarray*}
Letting $P$ be the estimator $\widehat{P}^{(-k)}$ and assuming  $\widehat{\lambda}^{(-k)}_1(X)$ is bounded away from zero with probability 1, 
it follows that
\begin{eqnarray*}
| Rem(\widehat{P}^{(-k)},P^*)  | & \leq & C^{(-k)}_{1} \underbrace{ \norm{  \widehat{\lambda}^{(-k)}_1(X)  - \lambda^*_1(X) } _{L_2} }_{O_{P^*}(n_k^{-2/5})} \underbrace{ \norm{ \widehat{\tau}^{(-k)}_t(1,X) -  \tau^*_t(1,X)  }_{L_2}}_{O_{P^*}(n_k^{-2/5})} \\
\end{eqnarray*}
where $C^{(-k)}_{1}$ is  $O_{P^*}(1)$. 
This implies that that Regularity Condition 2 holds and $\sqrt{n_k} R_{2k}$ is $o_{P^*}(1)$

In terms of $R_{1k}$, we note that it is equal to $\int  \left\{ \phi(\widehat{P}^{(-k)})(o) - \phi(P^*)(o) \right\} d( P^{(k)}_{n_k}  - P^*)(o)$, where $\phi(P^*)(o) = G_{\mu_t^*}^{(A1,A3)}(P^*)(o) + \mu_t^*$. Thus, $\sqrt{n_k} R_{1k}$ will be  \newline $O_{P^*} \left( \norm{\phi(\widehat{P}^{(-k)}) - \phi(P^*) }_{L_2} \right)$.  Using the triangle and Cauchy-Schwarz inequalities, it can be shown that 
\begin{eqnarray*}
\norm{ \phi(\widehat{P}^{(-k)}) - \phi(P^*) }_{L2} & \leq & D^{(-k)}_{1} \underbrace{\norm{\widehat{\lambda}^{(-k)}_1(X)  - \lambda^*_1(X)  }_{L_2}}_{o_P^*(1)} + 
 D^{(-k)}_{2} \underbrace{\norm{ \widehat{\tau}^{(-k)}_t(1,X) -  \tau^*_t(1,X)  }_{L_2}}_{o_P^*(1)} 
\end{eqnarray*}
for $D^{(-k)}_{1}$ and $D^{(-k)}_{2}$ are $O_{P^*}(1)$.  Thus, $\norm{ \phi(\widehat{P}^{(-k)}) - \phi(P^*) }_{L2}$ is $o_{P^*}(1)$  (i.e., Regularity Condition 1 holds) and $\sqrt{n_k} R_{1k}$ is $o_{P^*}(1)$

\subsection{Theorem 3}

It can be shown that
\begin{eqnarray*}
Rem(P,P^*) 
& = &   E^* \left[  \left( \frac{    \pi^*_t(X) }{  \pi_t(X) } -1 \right)  \left( \tau_t^*(X) -  \tau_t(X) \right)  \right] 
\end{eqnarray*}
Letting $P$ be the estimator $\widehat{P}^{(-k)}$ and assuming  $\widehat{\pi}^{(-k)}_t(X)$ is bounded away from zero with probability 1, 
it follows that
\begin{eqnarray*}
| Rem(\widehat{P}^{(-k)},P^*)  | & \leq & C^{(-k)}_{1} \underbrace{ \norm{  \widehat{\pi}^{(-k)}_t(X)  - \pi^*_t(X) } _{L_2} }_{O_{P^*}(n_k^{-2/5})} \underbrace{ \norm{ \widehat{\tau}^{(-k)}_t(X) -  \tau^*_t(X)  }_{L_2}}_{O_{P^*}(n_k^{-2/5})} \\
\end{eqnarray*}
where $C^{(-k)}_{1}$ is  $O_{P^*}(1)$. 
This implies that that Regularity Condition 2 holds and $\sqrt{n_k} R_{2k}$ is $o_{P^*}(1)$

In terms of $R_{1k}$, we note that it is equal to $\int  \left\{ \phi(\widehat{P}^{(-k)})(o) - \phi(P^*)(o) \right\} d( P^{(k)}_{n_k}  - P^*)(o)$, where $\phi(P^*)(o) = G_{\mu_t^*}^{(A1,A2,A3)}(P^*)(o) + \mu_t^*$. Thus, $\sqrt{n_k} R_{1k}$ will be  \newline $O_{P^*} \left( \norm{\phi(\widehat{P}^{(-k)}) - \phi(P^*) }_{L_2} \right)$.  Using the triangle and Cauchy-Schwarz inequalities, it can be shown that 
\begin{eqnarray*}
\norm{ \phi(\widehat{P}^{(-k)}) - \phi(P^*) }_{L2} & \leq & D^{(-k)}_{1} \underbrace{\norm{\widehat{\pi}^{(-k)}_t(X)  - \pi^*_t(X)  }_{L_2}}_{o_P^*(1)} + 
 D^{(-k)}_{2} \underbrace{\norm{ \widehat{\tau}^{(-k)}_t(X) -  \tau^*_t(X)  }_{L_2}}_{o_P^*(1)} 
\end{eqnarray*}
for $D^{(-k)}_{1}$ and $D^{(-k)}_{2}$ are $O_{P^*}(1)$.  Thus, $\norm{ \phi(\widehat{P}^{(-k)}) - \phi(P^*) }_{L2}$ is $o_{P^*}(1)$  (i.e., Regularity Condition 1 holds) and $\sqrt{n_k} R_{1k}$ is $o_{P^*}(1)$

\subsection{Theorem 4}

It can be shown that
\begin{eqnarray*}
Rem(P,P^*) & = &  \left(1 - \frac{\lambda_1^*}{\lambda_1} \right) ( \nu_t - \nu_t^*) 
\end{eqnarray*}
Letting $P$ be the estimator $\widehat{P}^{(-k)}$, 
it follows that
\begin{eqnarray*}
| Rem(\widehat{P}^{(-k)},P^*)  | & \leq & 
 C^{(-k)}_{1} \underbrace{ | \;   \widehat{\lambda}^{(-k)}_{1} - \lambda_{1}^* \; | }_{ O_{P^*}(1/\sqrt{n})} \underbrace{ | \widehat{\nu}_t^{(-k)} - \nu_t^* | }_{ o_{P^*}(1) }  
\end{eqnarray*}
where $C^{(-k)}_{1}$ is $O_{P^*}(1)$. 
This implies that that Regularity Condition 2 holds and $\sqrt{n_k} R_{2k}$ is $o_{P^*}(1)$

Using the triangle and Cauchy-Schwarz inequalities, it can be shown that 
\begin{eqnarray*}
\norm{ G(\widehat{P}^{(-k)}) - G(P^*) }_{L2} & \leq &  D^{(-k)}_{1} \underbrace{| \; \widehat{\lambda}^{(-k)}_{1} - \lambda^*_{1} \; |}_{o_P^*(1)} +  D^{(-k)}_{2} \underbrace{\norm{ \widehat{\tau}^{(-k)}_t(1,X) -  \tau^*_t(1,X)  }_{L_2}}_{o_P^*(1)} + \\
&& D^{(-k)}_{3} \underbrace{| \; \widehat{\nu}^{(-k)}_{t} - \nu^*_{t} \; |}_{o_P^*(1)}
\end{eqnarray*}
for $D^{(-k)}_{1}$, $D^{(-k)}_{2}$, and $D^{(-k)}_{3}$ are $O_{P^*}(1)$.  Thus, Regularity Condition 1 holds  and $\sqrt{n_k} R_{1k}$ is $o_{P^*}(1)$

\subsection{Theorem 5}

It can be shown that
\begin{eqnarray*}
Rem(P,P^*) & = &  \frac{1}{\lambda_1}   E^*\left[ \frac{ \lambda_1(X)} {\pi_t(X)}  \left( \pi_t^*(X) -\pi_t(X) \right) \left( \tau_t(X) -\tau^*_t(X) \right) \right] + \\
&& \frac{1}{\lambda_1}   E^*\left[ \left( \lambda_1(X) -\lambda_1^*(X) \right) \left( \tau_t(X) -\tau^*_t(X) \right) \right] + 
 \left(1 - \frac{\lambda_1^*}{\lambda_1} \right) ( \nu_t - \nu_t^*) 
\end{eqnarray*}
Letting $P$ be the estimator $\widehat{P}^{(-k)}$, 
it follows that
\begin{eqnarray*}
| Rem(\widehat{P}^{(-k)},P^*)  | & \leq & C^{(-k)}_{1} \underbrace{ \norm{  \widehat{\pi}_t^{(-k)}(X)  - \pi^*_t(X) } _{L_2} }_{O_{P^*}(n_k^{-2/5})} \underbrace{ \norm{ \widehat{\tau}^{(-k)}_t(X) -  \tau^*_t(X)  }_{L_2}}_{O_{P^*}(n_k^{-2/5})}  + \\
&& C^{(-k)}_{2} \underbrace{ \norm{  \widehat{\lambda}_1^{(-k)}(X)  - \lambda^*_1(X) } _{L_2} }_{O_{P^*}(n_k^{-2/5})} \underbrace{ \norm{ \widehat{\tau}^{(-k)}_t(X) -  \tau^*_t(X)  }_{L_2}}_{O_{P^*}(n_k^{-2/5})} + \\
& & C^{(-k)}_{3} \underbrace{ | \;   \widehat{\lambda}^{(-k)}_{1} - \lambda_{1}^* \; | }_{ O_{P^*}(n_k^{-1/2})} \underbrace{ | \widehat{\nu}_t^{(-k)} - \nu_t^* | }_{ o_{P^*}(1) }  
\end{eqnarray*}
where $C^{(-k)}_{1}$, $C^{(-k)}_{2}$ and  $C^{(-k)}_{3}$ are $O_{P^*}(1)$. 
This implies that that Regularity Condition 2 holds and $\sqrt{n_k} R_{2k}$ is $o_{P^*}(1)$

Using the triangle and Cauchy-Schwarz inequalities, it can be shown that 
\begin{eqnarray*}
\norm{ G(\widehat{P}^{(-k)}) - G(P^*) }_{L2} & \leq &  D^{(-k)}_{1} \underbrace{ \norm{ \widehat{\pi}^{(-k)}_{t}(X) - \pi^*_{t}(X) }_{L_2} }_{o_P^*(1)} + D^{(-k)}_{2} \underbrace{ \norm{ \widehat{\lambda}^{(-k)}_{1}(X) - \lambda^*_{1}(X) }_{L_2} }_{o_P^*(1)} + \\
&& 
D^{(-k)}_{3} \underbrace{\norm{ \widehat{\tau}^{(-k)}_t(X) -  \tau^*_t(X)  }_{L_2}}_{o_P^*(1)} + 
D^{(-k)}_{4} \underbrace{| \; \widehat{\lambda}^{(-k)}_{1} - \lambda^*_{1} \; |}_{o_P^*(1)} + \\
&& D^{(-k)}_{5} \underbrace{| \; \widehat{\nu}^{(-k)}_{t} - \nu^*_{t} \; |}_{o_P^*(1)}
\end{eqnarray*}
for $D^{(-k)}_{1}$, $D^{(-k)}_{2}$, $D^{(-k)}_{3}$, $D^{(-k)}_{4}$ and $D^{(-k)}_{5}$ that are $O_{P^*}(1)$.  Thus, Regularity Condition 1 holds  and $\sqrt{n_k} R_{1k}$ is $o_{P^*}(1)$

 \section{Simulation studies} \label{chap2:sec:simu}
 
We present simulation studies, motivated by the BARI study, to evaluate the performance of the five proposed estimators.   We consider three simulation studies, 
corresponding to the three sets of assumption represented in Figure 1 of the manuscript.  For each set of assumptions, we consider various model misspecification scenarios.  We simulated 5000 datasets, each with a sample size of $n=2000$.  Datasets were analyzed with $K=5$ sample splits. We report average bias, average of standard error estimate, Monte Carlo standard deviation of the estimator, coverage of 95\% Wald confidence intervals and Monte Carlo mean squared error for the three estimators of $\mu_1^*$ and $\mu_0^*$ and for the two estimators of $\nu_1^*$ and $\nu_0^*$.   Since we have assumed that (A1) holds for all scenarios and $\widetilde{\nu}_t^{(A1)}$ is immune to model misspecification, all simulations demonstrate that it is unbiased, with excellence correspondence between the average standard error and standard deviation of parameter estimates and coverage of 95\% confidence intervals close to their nominal level.

\subsection{Study 1: Assumptions (A1) and (A2)}

In this simulation study, we let $\lambda_1^*(X)$, $\pi_1^*(0,X)$ , $\tau_1^*(0,X)$ and $\tau_0^*(0,X)$, 
be equal to the estimated functions obtained from fitting the generalized additive logistic models to the BARI data. We generated data according to the following procedure:
\begin{enumerate}[1.]
\item Draw $X$ from the empirical distribution of the 12 covariates utilized in our analysis of the BARI study;
\item Draw $U \sim \mbox{Normal}(0,1)$ independent of $X$;
\item Draw $R \sim \mbox{Bernoulli}(\Phi(\Delta_R(X)+U);0,1)$ where $\Delta_R(X) = \Phi^{-1}(  \lambda^*_1(X),0,2)$ and $\Phi(\cdot,a,b)$ is the cumulative distribution function of a normal random variable with mean $a$ and variance $b$;
\item If $R=1$, draw $T \sim  \mbox{Bernoulli}(0.5)$;  If $R=0$, draw $T \sim \mbox{Bernoulli}(\pi_1^*(0,X))$
\item Let $\Delta_{Y_t}(X)$ be the solution to 
\[
\Phi_2(\Delta_{Y_t}(X),\Delta_R(X);\bold{0},\bold{\Sigma}) = \tau_1^*(X) \{ 1- \lambda_1^*(X) \}
\]
where $\Phi_2(\cdot,\cdot;\bold{a},\bold{B})$ is the cumulative distribution function of a bivariate normal random vector mean $\bold{a}$ and variance-covariance matrix $\bold{B}$ and $\bold{\Sigma}$ is a variance-covariance matrix with variance 1 and covariance 2.  For $t=0,1$, draw $Y_t  \sim \mbox{Bernoulli}(\Phi(\Delta_{Y_t}(X) + U;0,1))$
\item Set $Y = T Y_1 + (1-T)Y_0$
\end{enumerate}
Steps 3 and 5 use the idea of generating correlated binary outcomes discussed by \cite{swihart2014unifying}; the conditional distribution of $R$ given $X$ is  $\mbox{Bernoulli}(\lambda_1^*(X))$ and the conditional distribution of $Y_t$ given $R=0$ and $X$ is  $\mbox{Bernoulli}(\tau^*_t(0,X))$. 

Table \ref{a1a2}  presents the results of the simulation study.  In addition to evaluating the five estimators under correct model specification - (a), we considered the following three misspecification scenarios:
\begin{enumerate}[(a)]
\setcounter{enumi}{1}
\item $\widehat{\pi}_1(0,X)$ is replaced by $\Phi(\mbox{logit} \{ \widehat{\pi}_1(0,X) \}; 0, 25)$;
\item $\widehat{\tau}_t(0,X)$ is replaced by $\Phi(\mbox{logit} \{\widehat{\tau}_t(0,X) \}; 0, 25)$; 
\item $\widehat{\pi}_1(0,X)$ and $\widehat{\tau}_t(0,X)$ are replaced by $\Phi(\mbox{logit} \{ \widehat{\pi}_1(0,X) \}; 0, 25)$ and \\ $\Phi(\mbox{logit} \{\widehat{\tau}_t(0,X) \}; 0, 25)$, respectively.
\end{enumerate}

Table \ref{a1a2} shows that, when {\em either} the model for $\pi^*_t(0,X)$ or $\tau_t^*(X)$ is modeled correctly, then $\widetilde{\mu}_t^{(A1,A2)}$ performs well, but it performs poorly when both models are misspecified.  As expected, the estimators $\widetilde{\mu}_t^{(A1,A3)}$, $\widetilde{\mu}_t^{(A1,A2,A3)}$ and  $\widetilde{\nu}_t^{(A1,A2,A3)}$ perform poorly for all scenarios.  

\begin{table}
\footnotesize
\centering
\caption{Simulation results: (A1), (A2)} \label{a1a2}
\begin{tabular}{llccccc}
  \multicolumn{7}{c}{(a)  $\pi_t^*(0,X)$ and $\tau_t^*(0,X)$ modeled correctly.} \\ \hline
 Parameter&Estimator& {Bias} & Mean SE & SD & \parbox{1cm}{\centering 95\% CI \\ coverage} & $\sqrt{\mbox{MSE}}$ \\
  \hline
  $\mu_1^*$ & $\widetilde{\mu}_1^{(A1,A2)}$ & 0.03\% & 1.23\% & 1.23\% & 94.72\% &  1.23\%\\  
 &$\widetilde{\mu}_1^{(A1,A3)}$  & 10.09\% & 2.09\% & 2.09\% & 0.10\% & 10.30\%\\ 
 &$\widetilde{\mu}_1^{(A1,A2,A3)}$  & -1.13\% & 1.18\% & 1.17\% & 83.04\% & 1.63\%\\ 
  \hline
  $\mu_0^*$ & $\widetilde{\mu}_0^{(A1,A2)}$& -0.01\% & 1.36\% & 1.34\% & 95.30\% & 1.34\% \\ 
 & $\widetilde{\mu}_0^{(A1,A3)}$ & 8.78\% & 1.96\% & 1.94\% & 0.40\% & 8.99\%\\ 
 & $\widetilde{\mu}_0^{(A1,A2,A3)}$ & 1.58\% & 1.36\% & 1.36\% & 79.86\% & 2.08\%\\ 
  \hline
 \hline
$\nu_{1}^*$ & $\widetilde{\nu}_1^{(A1)}$ & 0.04\% & 2.03\% & 2.05\% & 94.82\% & 2.05\% \\ 
 &$\widetilde{\nu}_1^{(A1,A2,A3)}$& -11.18\% & 1.32\% & 1.31\% & 0.00\% &  11.25\%\\ 
   \hline
  $\nu_{0}^*$ & $\widetilde{\nu}_0^{(A1)}$ & 0.01\% & 1.94\% & 1.95\% & 95.06\% &  1.95\%\\ 
 & $\widetilde{\nu}_0^{(A1,A2,A3)}$& -7.24\% & 1.49\% & 1.50\% & 0.32\% & 7.39\% \\
  \hline 
  \hlinewd{1pt}\\[-1em]
     \multicolumn{7}{c}{(b) $\pi_t^*(0,X)$ modeled incorrectly, $\tau_t^*(0,X)$ modeled correctly} \\ \hline
Parameter&Estimator& {Bias} & Mean SE & SD & \parbox{1cm}{\centering 95\% CI \\ coverage} & $\sqrt{\mbox{MSE}}$ \\
  \hline
  $\mu_1^*$ & $\widetilde{\mu}_1^{(A1,A2)}$ & 0.00\% & 1.23\% & 1.18\% & 95.60\% &  1.18\%\\ 
 &$\widetilde{\mu}_1^{(A1,A3)}$ & 9.66\% & 2.09\% & 2.07\% & 0.16\% &  9.88\%\\ 
 &$\widetilde{\mu}_1^{(A1,A2,A3)}$ & -1.07\% & 1.22\% & 1.14\% & 86.80\% &  1.56\%\\ 
  \hline
  $\mu_0^*$ & $\widetilde{\mu}_0^{(A1,A2)}$& 0.06\% & 1.16\% & 1.26\% & 93.00\% & 1.26\%\\  
 & $\widetilde{\mu}_0^{(A1,A3)}$ & 8.35\% & 1.95\% & 1.97\% & 0.78\% &  8.58\%\\ 
 & $\widetilde{\mu}_0^{(A1,A2,A3)}$ & 1.53\% & 1.19\% & 1.33\% & 72.64\% &  2.03\%\\ 
  \hline
 \hline
$\nu_{1}^*$ & $\widetilde{\nu}_1^{(A1)}$ & -0.01\% & 2.02\% & 2.03\% & 94.38\% &  2.03\%\\ 
 &$\widetilde{\nu}_1^{(A1,A2,A3)}$& -10.56\% & 1.36\% & 1.28\% & 0.00\% &  10.64\%\\ 
   \hline
  $\nu_{0}^*$ & $\widetilde{\nu}_0^{(A1)}$ & 0.00\% & 1.92\% & 1.96\% & 94.88\% & 1.96\% \\ 
 & $\widetilde{\nu}_0^{(A1,A2,A3)}$& -6.70\% & 1.36\% & 1.47\% & 0.40\% &  6.86\%\\ 
  \hline  
  \hlinewd{1pt}\\[-1em]
     \multicolumn{7}{c}{(c) $\pi_t^*(0,X)$ modeled correctly, $\tau_t^*(0,X)$ modeled incorrectly} \\ \hline
 Parameter&Estimator& {Bias} & Mean SE & SD & \parbox{1cm}{\centering 95\% CI \\ coverage} & $\sqrt{\mbox{MSE}}$ \\
  \hline
  $\mu_1^*$ & $\widetilde{\mu}_1^{(A1,A2)}$ & -0.18\% & 1.23\%& 1.18\%& 95.38\% & 1.19\% \\ 
 &$\widetilde{\mu}_1^{(A1,A3)}$ & 9.93\%& 2.09\% & 2.02\% & 0.04\% & 10.13\%\\ 
&$\widetilde{\mu}_1^{(A1,A2,A3)}$ & -1.11\% & 1.16\% & 1.12\% & 83.60\% &  1.58\%\\ 
  \hline
  $\mu_0^*$ & $\widetilde{\mu}_0^{(A1,A2)}$& -0.51\% & 1.42\%& 1.35\% & 94.36\% & 1.44\% \\ 
 & $\widetilde{\mu}_0^{(A1,A3)}$ & 8.65\% & 1.94\% & 1.95\% & 0.32\% & 8.87\% \\ 
& $\widetilde{\mu}_0^{(A1,A2,A3)}$ & 1.48\% & 1.33\% & 1.35\% & 81.08\% &  2.00\%\\ 
   \hline
 \hline
$\nu_{1}^*$ & $\widetilde{\nu}_1^{(A1)}$ & 0.02\%& 2.02\%& 1.98\%& 95.30\%& 1.98\% \\ 
 &$\widetilde{\nu}_1^{(A1,A2,A3)}$& -10.93\% & 1.28\%& 1.25\% & 0.00\% & 11.00\% \\ 
   \hline
  $\nu_{0}^*$ & $\widetilde{\nu}_0^{(A1)}$ & -0.03\% & 1.92\% & 1.95\% & 94.14\%& 1.95\% \\ 
 & $\widetilde{\nu}_0^{(A1,A2,A3)}$& -7.16\% & 1.46\% & 1.47\%& 0.32\%& 7.31\% \\ 
  \hline
   \hlinewd{1pt}\\[-1em]
     \multicolumn{7}{c}{(d) $\pi_t^*(0,X)$ and $\tau_t^*(0,X)$ modeled incorrectly} \\ \hline
 Parameter&Estimator& {Bias} & Mean SE & SD & \parbox{1cm}{\centering 95\% CI \\ coverage} & $\sqrt{\mbox{MSE}}$ \\
  \hline
  $\mu_1^*$ & $\widetilde{\mu}_1^{(A1,A2)}$ & -1.68\% & 1.28\%& 1.21\%& 74.32\% & 2.07\% \\ 
 &$\widetilde{\mu}_1^{(A1,A3)}$ & 9.75\%& 2.08\% & 2.03\% & 0.14\% & 9.96\%\\ 
&$\widetilde{\mu}_1^{(A1,A2,A3)}$ & -1.18\% & 1.23\% & 1.15\% & 84.50\% &  1.65\%\\
  \hline
  $\mu_0^*$ & $\widetilde{\mu}_0^{(A1,A2)}$& 1.79\% & 1.19\%& 1.36\% & 66.26\% & 2.25\% \\ 
 & $\widetilde{\mu}_0^{(A1,A3)}$ & 8.46\% & 1.94\% & 1.91\% & 0.32\% & 8.67\% \\ 
& $\widetilde{\mu}_0^{(A1,A2,A3)}$ & 1.61\% & 1.20\% & 1.33\% & 72.62\% &  2.09\%\\ 
   \hline
 \hline
$\nu_{1}^*$ & $\widetilde{\nu}_1^{(A1)}$ & -0.05\%& 2.02\%& 2.00\%& 95.24\%& 2.01\% \\ 
 &$\widetilde{\nu}_1^{(A1,A2,A3)}$& -10.97\% & 1.37\%& 1.29\% & 0.00\% & 11.05\% \\ 
   \hline
  $\nu_{0}^*$ & $\widetilde{\nu}_0^{(A1)}$ & 0.03\% & 1.93\% & 1.94\% & 94.66\%& 1.94\% \\ 
 & $\widetilde{\nu}_0^{(A1,A2,A3)}$& -6.84\% & 1.37\% & 1.48\%& 0.46\%& 7.00\% \\ 
  \hline
  \hlinewd{1pt}
\end{tabular}
\end{table}

\subsection{Study 2: Assumptions (A1) and (A3)}

In this simulation study, we let $\lambda_1^*(X)$, $\pi_1^*(0,X)$, $\tau_1^*(1,X)$ and $\tau_0^*(1,X)$, 
be equal to the estimated functions obtained from fitting the generalized additive logistic models to the BARI data. We generated data according to the following procedure:
\begin{enumerate}[1.]
\item Draw $X$ from the empirical distribution of the 12 covariates utilized in our analysis of the BARI study;
\item Draw $U \sim \mbox{Normal}(0,1)$ independent of $X$;
\item Draw $R \sim \mbox{Bernoulli}(\lambda_1^*(X))$ 
\item If $R=1$, draw $T \sim  \mbox{Bernoulli}(0.5)$;  If $R=0$, draw $T \sim \mbox{Bernoulli}(\mbox{expit} \{ \mbox{logit} \{ \pi_1^*(0,X)) - 2 U \} \})$
\item Draw $Y_t \sim \mbox{Bernoulli}(\Phi(\Delta_{Y_t}(X)+U);0,1)$, where $\Delta_{Y_t}(X) = \Phi^{-1}(  \tau^*_t(1,X),0,2)$.\item Set $Y = T Y_1 + (1-T)Y_0$
\end{enumerate}
Step 5 ensures that the conditional distribution of $Y_t$ given $X$ is  $\mbox{Bernoulli}( \tau^*_t(1,X))$.

Table \ref{a1a3}  presents the results of the simulation study.  In addition to evaluating the five estimators under correct model specification - (a), we considered the following three misspecification scenarios:
\begin{enumerate}[(a)]
\setcounter{enumi}{1}
\item $\widehat{\lambda}_1(X)$ is replaced by $ 0.7 \Phi(\mbox{logit} \{ \widehat{\lambda}_1(X) \}; 0, 25) + 0.3 \Phi(\mbox{logit} \{ \widehat{\lambda}_1(X) \}; 0.8, 0.04)$; 
\item $\widehat{\tau}_t(1,X)$ is replaced by $\Phi(\mbox{logit} \{\widehat{\tau}_t(1,X) \}; 0, 25)$; 
\item $\widehat{\lambda}_1(X)$ and $\widehat{\tau}_t(1,X)$ are replaced by $ 0.7 \Phi(\mbox{logit} \{ \widehat{\lambda}_1(X) \}; 0, 25) + \\ 0.3 \Phi(\mbox{logit} \{ \widehat{\lambda}_1(X) \}; 0.8, 0.04)$ and $\Phi(\mbox{logit} \{\widehat{\tau}_t(1,X) \}; 0, 25)$, respectively.
\end{enumerate}

Table \ref{a1a3} shows that, when {\em either} the model for $\lambda_1^*(X)$ or $\tau_t^*(1,X)$ is modeled correctly, then $\widetilde{\mu}_t^{(A1,A3)}$ performs well, but it performs poorly when both models are misspecified.  As expected, the estimators $\widetilde{\mu}_t^{(A1,A2)}$, $\widetilde{\mu}_t^{(A1,A2,A3)}$ and  $\widetilde{\nu}_t^{(A1,A2,A3)}$ perform poorly for all scenarios.

\begin{table}
\footnotesize
\centering
\caption{Simulation results: (A1), (A3)} \label{a1a3}
\begin{tabular}{llccccc}
  \multicolumn{7}{c}{(a) $\lambda_1^*(X)$, $\tau_t^*(1,X)$ modeled correctly} \\ \hline
 Parameter&Estimator& {Bias} & Mean SE & SD & \parbox{1cm}{\centering 95\% CI \\ coverage} & $\sqrt{\mbox{MSE}}$ \\
  \hline
  $\mu_1^*$ & $\widetilde{\mu}_1^{(A1,A2)}$ & -3.25\% & 0.91\% & 0.90\% & 6.72\% &  3.37\%\\  
 &$\widetilde{\mu}_1^{(A1,A3)}$  & -0.02\% & 1.53\% & 1.51\% & 94.84\% & 1.51\%\\ 
 &$\widetilde{\mu}_1^{(A1,A2,A3)}$  & -3.50\% & 0.88\% & 0.86\% & 2.98\% & 3.60\%\\ 
  \hline
  $\mu_0^*$ & $\widetilde{\mu}_0^{(A1,A2)}$&  4.25\% & 1.29\% & 1.27\% & 6.90\% & 4.43\%\\ 
 & $\widetilde{\mu}_0^{(A1,A3)}$ & -0.01\% & 1.36\% & 1.34\% & 95.02\% & 1.34\%\\ 
 & $\widetilde{\mu}_0^{(A1,A2,A3)}$ & 3.64\% & 1.14\% & 1.14\% & 9.04\% & 3.82\%\\ 
  \hline
 \hline
$\nu_{1}^*$ & $\widetilde{\nu}_1^{(A1)}$ & -0.02\% & 1.47\% & 1.47\% & 94.88\% & 1.47\% \\ 
 &$\widetilde{\nu}_1^{(A1,A2,A3)}$& -3.43\% & 0.97\% & 0.94\% & 7.02\% &  3.55\%\\ 
   \hline
  $\nu_{0}^*$ & $\widetilde{\nu}_0^{(A1)}$ & -0.01\% & 1.32\% & 1.33\% & 94.62\% &  1.33\%\\ 
 & $\widetilde{\nu}_0^{(A1,A2,A3)}$& 3.58\% & 1.18\% & 1.19\% & 12.24\% & 3.77\% \\
  \hline  
  \hlinewd{1pt}\\[-1em]
     \multicolumn{7}{c}{(b) $\lambda_1^*(X)$ modeled incorrectly, $\tau_t^*(1,X)$ modeled correctly} \\ \hline
Parameter&Estimator& {Bias} & Mean SE & SD & \parbox{1cm}{\centering 95\% CI \\ coverage} & $\sqrt{\mbox{MSE}}$ \\
  \hline
  $\mu_1^*$ & $\widetilde{\mu}_1^{(A1,A2)}$ & -3.30\% & 0.91\% & 0.91\% & 6.12\% &  3.43\%\\ 
 &$\widetilde{\mu}_1^{(A1,A3)}$ & -0.05\% & 2.00\% & 1.52\% & 98.88\% &  1.52\%\\ 
 &$\widetilde{\mu}_1^{(A1,A2,A3)}$ & -3.52\% & 0.87\% & 0.88\% & 2.94\% &  3.62\%\\ 
  \hline
  $\mu_0^*$ & $\widetilde{\mu}_0^{(A1,A2)}$& 4.08\% & 1.28\% & 1.23\% & 7.66\% & 4.27\%\\  
 & $\widetilde{\mu}_0^{(A1,A3)}$ & -0.04\% & 1.80\% & 1.37\% & 98.66\% &  1.37\%\\ 
 & $\widetilde{\mu}_0^{(A1,A2,A3)}$ & 3.54\% & 1.18\% & 1.12\% & 11.74\% &  3.71\%\\ 
  \hline
 \hline
$\nu_{1}^*$ & $\widetilde{\nu}_1^{(A1)}$ & -0.01\% & 1.47\% & 1.48\% & 94.64\% &  1.48\%\\ 
 &$\widetilde{\nu}_1^{(A1,A2,A3)}$& -3.39\% & 0.74\% & 0.96\% & 3.44\% &  3.52\%\\ 
   \hline
  $\nu_{0}^*$ & $\widetilde{\nu}_0^{(A1)}$ & -0.01\% & 1.31\% & 1.33\% & 94.60\% & 1.33\% \\ 
 & $\widetilde{\nu}_0^{(A1,A2,A3)}$& 3.46\% & 0.92\% & 1.15\% & 6.08\% &  3.65\%\\ 
  \hline  
  \hlinewd{1pt}\\[-1em]
    \multicolumn{7}{c}{(c)  $\lambda_1^*(X)$ modeled correctly, $\tau_t^*(1,X)$ modeled incorrectly} \\ \hline
 Parameter&Estimator& {Bias} & Mean SE & SD & \parbox{1cm}{\centering 95\% CI \\ coverage} & $\sqrt{\mbox{MSE}}$ \\
  \hline
  $\mu_1^*$ & $\widetilde{\mu}_1^{(A1,A2)}$ & -3.33\% & 0.96\%& 0.96\%& 7.28\% & 3.47\% \\ 
 &$\widetilde{\mu}_1^{(A1,A3)}$ & -0.24\%& 1.71\% & 1.64\% & 95.16\% & 1.66\%\\ 
&$\widetilde{\mu}_1^{(A1,A2,A3)}$ & -3.55\% & 0.88\% & 0.89\% & 3.22\% &  3.66\%\\ 
  \hline
  $\mu_0^*$ & $\widetilde{\mu}_0^{(A1,A2)}$& 4.25\% & 1.34\%& 1.31\% & 9.10\% & 4.44\% \\ 
 & $\widetilde{\mu}_0^{(A1,A3)}$ & -0.22\% & 1.58\% & 1.50\% & 95.10\% & 1.52\% \\ 
& $\widetilde{\mu}_0^{(A1,A2,A3)}$ & 3.67\% & 1.14\% & 1.14\% & 8.38\% &  3.84\%\\ 
   \hline
 \hline
$\nu_{1}^*$ & $\widetilde{\nu}_1^{(A1)}$ & 0.00\%& 1.58\%& 1.59\%& 94.28\%& 1.59\% \\ 
 &$\widetilde{\nu}_1^{(A1,A2,A3)}$& -3.43\% & 0.97\%& 0.97\% & 7.60\% & 3.56\% \\ 
   \hline
  $\nu_{0}^*$ & $\widetilde{\nu}_0^{(A1)}$ & 0.03\% & 1.46\% & 1.47\% & 94.42\%& 1.47\% \\ 
 & $\widetilde{\nu}_0^{(A1,A2,A3)}$& 3.57\% & 1.17\% & 1.17\%& 11.88\%& 3.76\% \\ 
  \hline
   \hlinewd{1pt}\\[-1em]
     \multicolumn{7}{c}{(d)  $\lambda_1^*(X)$, $\tau_t^*(1,X)$ modeled incorrectly} \\ \hline
 Parameter&Estimator& {Bias} & Mean SE & SD & \parbox{1cm}{\centering 95\% CI \\ coverage} & $\sqrt{\mbox{MSE}}$ \\
  \hline
  $\mu_1^*$ & $\widetilde{\mu}_1^{(A1,A2)}$ & -3.22\% & 0.96\%& 0.94\%& 8.68\% & 3.36\% \\ 
 &$\widetilde{\mu}_1^{(A1,A3)}$ & -6.76\%& 2.24\% & 1.85\% & 10.92\% & 7.01\%\\ 
&$\widetilde{\mu}_1^{(A1,A2,A3)}$ & -3.44\% & 0.88\% & 0.87\% & 3.20\% &  3.55\%\\ 
  \hline
  $\mu_0^*$ & $\widetilde{\mu}_0^{(A1,A2)}$& 4.08\% & 1.32\%& 1.27\% & 10.02\% & 4.27\% \\ 
 & $\widetilde{\mu}_0^{(A1,A3)}$ & -7.23\% & 2.08\% & 1.67\% & 3.90\% & 7.42\% \\ 
& $\widetilde{\mu}_0^{(A1,A2,A3)}$ & 3.54\% & 1.18\% & 1.12\% & 11.26\% &  3.71\%\\ 
   \hline
 \hline
$\nu_{1}^*$ & $\widetilde{\nu}_1^{(A1)}$ & -0.00\%& 1.58\%& 1.59\%& 94.62\%& 1.59\% \\ 
 &$\widetilde{\nu}_1^{(A1,A2,A3)}$& -3.38\% & 0.75\%& 0.96\% & 3.60\% & 3.51\% \\ 
   \hline
  $\nu_{0}^*$ & $\widetilde{\nu}_0^{(A1)}$ & -0.04\% & 1.46\% & 1.47\% & 94.36\%& 1.47\% \\ 
 & $\widetilde{\nu}_0^{(A1,A2,A3)}$& 3.47\% & 0.93\% & 1.16\%& 5.86\%& 3.66\% \\ 
  \hline
  \hlinewd{1pt}
\end{tabular}
\end{table}

\subsection{Study 3: Assumptions (A1), (A2) and (A3)}

In this simulation study, we let $\lambda_1^*(X)$, $\pi_1^*(0,X)$, $\tau_1^*(X)$ and $\tau_0^*(X)$, 
be equal to the estimated functions obtained from fitting the generalized additive logistic models to the BARI data. We generated data according to the following procedure:
\begin{enumerate}[1.]
\item Draw $X$ from the empirical distribution of the 12 covariates utilized in our analysis of the BARI study;
\item Draw $U \sim \mbox{Normal}(0,1)$ independent of $X$;
\item Draw $R \sim \mbox{Bernoulli}(\lambda_1^*(X))$ 
\item If $R=1$, draw $T \sim  \mbox{Bernoulli}(0.5)$;  If $R=0$, draw $T \sim \mbox{Bernoulli}(\pi_1^*(0,X)))$
\item Draw $Y_t \sim \mbox{Bernoulli}(\Phi(\Delta_{Y_t}(X)+U);0,1)$, where $\Delta_{Y_t}(X) = \Phi^{-1}(  \tau^*_t(X),0,2)$.
\item Set $Y = T Y_1 + (1-T)Y_0$
\end{enumerate}
Step 5 ensures that the conditional distribution of $Y_t$ given $X$ is  $\mbox{Bernoulli}( \tau^*_t(X))$.

Table \ref{a1a2a3}  presents the results of the simulation study.  In addition to evaluating the five estimators under correct model specification - (a), we considered the following seven misspecification scenarios:
\begin{enumerate}[(a)]
\setcounter{enumi}{1}
\item $\widehat{\lambda}_1(X)$ is replaced by $ 0.7 \Phi(\mbox{logit} \{ \widehat{\lambda}_1(X) \}; 0, 25) + 0.3 \Phi(\mbox{logit} \{ \widehat{\lambda}_1(X) \}; 0.8, 0.04)$;
\item $\widehat{\pi}_1(0,X)$ is replaced by $\Phi(\mbox{logit} \{ \widehat{\pi}_1(0,X) \}; 0, 25)$;
\item $\widehat{\tau}_t(X)$, $\widehat{\tau}_t(1,X)$ and $\widehat{\tau}_t(0,X)$ are replaced by $\Phi(\mbox{logit} \{\widehat{\tau}_t(X) \}; 0, 25)$, $\Phi(\mbox{logit} \{\widehat{\tau}_t(1,X) \}; 0, 25)$ and \newline $\Phi(\mbox{logit} \{\widehat{\tau}_t(0,X) \}; 0, 25)$, respectively;
\item $\widehat{\lambda}_1(X)$ and $\widehat{\pi}_1(0,X)$ are replaced by $ 0.7 \Phi(\mbox{logit} \{ \widehat{\lambda}_1(X) \}; 0, 25) + 0.3 \Phi(\mbox{logit} \{ \widehat{\lambda}_1(X) \}; 0.8, 0.04)$ \newline and $\Phi(\mbox{logit} \{ \widehat{\pi}_1(1,X) \}; 0, 25)$, respectively;
\item $\widehat{\lambda}_1(X)$, $\widehat{\tau}_t(X)$, $\widehat{\tau}_t(1,X)$ and $\widehat{\tau}_t(0,X)$ are replaced by $ 0.7 \Phi(\mbox{logit} \{ \widehat{\lambda}_1(X) \}; 0, 25) + \\ 0.3 \Phi(\mbox{logit} \{ \widehat{\lambda}_1(X) \}; 0.8, 0.04)$, $\Phi(\mbox{logit} \{\widehat{\tau}_t(X) \}; 0, 25)$, $\Phi(\mbox{logit} \{\widehat{\tau}_t(1,X) \}; 0, 25)$ and \\$\Phi(\mbox{logit} \{\widehat{\tau}_t(0,X) \}; 0, 25)$, respectively;
\item $\widehat{\pi}_1(0,X)$, $\widehat{\tau}_t(X)$, $\widehat{\tau}_t(1,X)$ and $\widehat{\tau}_t(0,X)$ are replaced are replaced by $\Phi(\mbox{logit} \{ \widehat{\pi}_1(0,X) \}; 0, 25)$,  $\newline \Phi(\mbox{logit} \{\widehat{\tau}_t(X) \}; 0, 25)$, $\Phi(\mbox{logit} \{\widehat{\tau}_t(1,X) \}; 0, 25)$ and $\Phi(\mbox{logit} \{\widehat{\tau}_t(0,X) \}; 0, 25)$, respectively;
\item  $\widehat{\lambda}_1(X)$,  $\widehat{\pi}_1(0,X)$,  $\widehat{\tau}_t(X)$, $\widehat{\tau}_t(1,X)$ and $\widehat{\tau}_t(0,X)$  are replaced by \newline $ 0.7 \Phi(\mbox{logit} \{ \widehat{\lambda}_1(X) \}; 0, 25) + 0.3 \Phi(\mbox{logit} \{ \widehat{\lambda}_1(X) \}; 0.8, 0.04)$, \newline $\Phi(\mbox{logit} \{ \widehat{\pi}_1(0,X) \}; 0, 25)$, $\Phi(\mbox{logit} \{\widehat{\tau}_t(X) \}; 0, 25)$, \\ $\Phi(\mbox{logit} \{\widehat{\tau}_t(1,X) \}; 0, 25)$ and $\Phi(\mbox{logit} \{\widehat{\tau}_t(0,X) \}; 0, 25)$, respectively.
\end{enumerate}

Table \ref{a1a2a3} shows that, when {\em either} the model for $\lambda_1^*(X)$ or $\tau_t^*(1,X)$ is modeled correctly, then $\widetilde{\mu}_t^{(A1,A3)}$ performs well, but it performs poorly when both models are misspecified.  As expected, the estimators $\widetilde{\mu}_t^{(A1,A2)}$, $\widetilde{\mu}_t^{(A1,A2,A3)}$ and  $\widetilde{\nu}_t^{(A1,A2,A3)}$ perform poorly for all scenarios.  

Under simulation scenarios (a)-(e), we see that all estimators are unbiased.  Among the CC estimators,  $\widetilde{\mu}_t^{(A1,A2,A3)}$ is slightly more efficient than $\widetilde{\mu}_t^{(A1,A3)}$ and both are much more efficient than $\widetilde{\mu}_t^{(A1,A3)}$.   Between the two RCT estimators, $\widetilde{\nu}_t^{(A1,A2,A3)}$ is substantially more efficient than $\widetilde{\nu}_t^{(A1)}$.  As expected under model misspecification, we do see (1) some poor correspondence between the average standard error and the standard deviation of parameter estimates and (2) coverage of estimated 95\% confidence intervals different than their nominal level. For Scenario (b), this can seen for $\widetilde{\mu}_t^{(A1,A3)}$, $\widetilde{\mu}_t^{(A1,A2,A3)}$, and $\widetilde{\nu}_t^{(A1,A2,A3)}$.   For Scenario (c), it can also be seen for $\widetilde{\mu}_0^{(A1,A3)}$, $\widetilde{\mu}_0^{(A1,A2,A3)}$, and $\widetilde{\nu}_0^{(A1,A2,A3)}$. 

Under simulation scenarios (f)-(h),  the estimators that rely on Assumptions (A1), (A2) and (A3) perform poorly.   Under scenarios (f) and (h), estimators that rely on Assumptions (A1) and (A3) perform poorly, whereas for scenarios (g) and (h), estimators that rely Assumptions (A1) and (A2) that perform poorly. 

\begin{table}
\footnotesize
\centering
\caption{Simulation results: (A1), (A2), (A3)} \label{a1a2a3}
\begin{tabular}{llccccc}
  \multicolumn{7}{c}{(a) $\lambda_1^*(X)$, $\pi_t^*(0,X)$, $\tau_t^*(X)=\tau_t^*(1,X)=\tau_t^*(0,X)$ modeled correctly } \\ \hline
 Parameter&Estimator& {Bias} & Mean SE & SD & \parbox{2cm}{\centering 95\% CI \\ coverage} & $\sqrt{\mbox{MSE}}$ \\
  \hline
  $\mu_1^*$ & $\widetilde{\mu}_1^{(A1,A2)}$ & 0.01\% & 0.97\% & 0.94\% & 95.32\% &  0.94\%\\  
 &$\widetilde{\mu}_1^{(A1,A3)}$  & -0.02\% & 1.42\% & 1.39\% & 95.08\% & 1.39\%\\ 
 &$\widetilde{\mu}_1^{(A1,A2,A3)}$  & 0.01\% & 0.94\% & 0.91\% & 95.24\% & 0.91\%\\ 
  \hline
  $\mu_0^*$ & $\widetilde{\mu}_0^{(A1,A2)}$& -0.05\% & 1.20\% & 1.19\% & 95.02\% & 1.19\% \\ 
 & $\widetilde{\mu}_0^{(A1,A3)}$ & -0.01\% & 1.36\% & 1.34\% & 94.82\% & 1.34\%\\ 
 & $\widetilde{\mu}_0^{(A1,A2,A3)}$ & -0.01\% & 1.02\% & 1.02\% & 94.82\% & 1.02\%\\ 
  \hline
 \hline
$\nu_{1}^*$ & $\widetilde{\nu}_1^{(A1)}$ & 0.00\% & 1.40\% & 1.38\% & 95.22\% & 1.38\% \\ 
 &$\widetilde{\nu}_1^{(A1,A2,A3)}$& 0.02\% & 1.03\% & 1.00\% & 95.40\% &  1.00\%\\ 
   \hline
  $\nu_{0}^*$ & $\widetilde{\nu}_0^{(A1)}$ & 0.00\% & 1.34\% & 1.35\% & 94.68\% &  1.35\%\\ 
 & $\widetilde{\nu}_0^{(A1,A2,A3)}$& -0.00\% & 1.09\% & 1.10\% & 95.40\% & 1.00\% \\
  \hline   
  \hlinewd{1pt}\\[-1em]
  \multicolumn{7}{c}{(b) $\lambda_1^*(X)$ modeled incorrectly, $\pi_t^*(0,X)$, $\tau_t^*(X)=\tau_t^*(1,X)=\tau_t^*(0,X)$ modeled correctly} \\ \hline
 Parameter&Estimator& {Bias} & Mean SE & SD & \parbox{2cm}{\centering 95\% CI \\ coverage} & $\sqrt{\mbox{MSE}}$ \\
  \hline
  $\mu_1^*$ & $\widetilde{\mu}_1^{(A1,A2)}$  & -0.01\%  & 0.96\%  & 0.98\%  & 94.50\%  & 0.98\%  \\ 
 &$\widetilde{\mu}_1^{(A1,A3)}$ & -0.04\%  & 1.90\%  & 1.44\%  & 98.50\%  & 1.44\%  \\  
 &$\widetilde{\mu}_1^{(A1,A2,A3)}$ & -0.00\%  & 0.92\%  & 0.95\%  & 93.94\%  & 0.95\%  \\  
  \hline
  $\mu_0^*$ & $\widetilde{\mu}_0^{(A1,A2)}$ & -0.04\%  & 1.20\%  & 1.19\%  & 95.64\%  & 1.19\%  \\ 
 & $\widetilde{\mu}_0^{(A1,A3)}$ & -0.03\% & 1.80\%  & 1.35\%  & 98.84\%  & 1.35\%  \\ 
 & $\widetilde{\mu}_0^{(A1,A2,A3)}$ & -0.00\% & 1.09\%  & 1.02\%  & 95.80\%  & 1.02\%  \\ 
  \hline
 \hline
$\nu_{1}^*$ & $\widetilde{\nu}_1^{(A1)}$ & -0.02\%  & 1.40\%  & 1.43\% & 94.50\% & 1.43\%  \\ 
 &$\widetilde{\nu}_1^{(A1,A2,A3)}$& -0.01\%  & 0.77\%  & 1.03\%  & 85.48\%  & 1.03\%  \\
   \hline
  $\nu_{0}^*$ & $\widetilde{\nu}_0^{(A1)}$ & -0.01\%  & 1.33\% & 1.34\%  & 94.20\%  & 1.34\%  \\  
 & $\widetilde{\nu}_0^{(A1,A2,A3)}$& -0.01\%  & 0.88\% & 1.08\%  & 88.92\%  & 1.08\% \\ 
  \hline
   \hlinewd{1pt}
   \hlinewd{1pt}\\[-1em]  
    \multicolumn{7}{c}{(c) $\pi_t^*(0,X)$ modeled incorrectly, $\lambda_1^*(X)$, $\tau_t^*(X)=\tau_t^*(1,X)=\tau_t^*(0,X)$ modeled correctly} \\ \hline
Parameter&Estimator& {Bias} & Mean SE & SD & \parbox{2cm}{\centering 95\% CI \\ coverage} & $\sqrt{\mbox{MSE}}$ \\
  \hline
  $\mu_1^*$ & $\widetilde{\mu}_1^{(A1,A2)}$ & -0.02\% & 0.98\% & 0.95\% & 95.68\% &  0.95\%\\ 
 &$\widetilde{\mu}_1^{(A1,A3)}$ & 0.00\% & 1.42\% & 1.42\% & 94.58\% &  1.42\%\\ 
 &$\widetilde{\mu}_1^{(A1,A2,A3)}$ & -0.02\% & 0.98\% & 0.93\% & 96.18\% &  0.93\%\\ 
  \hline
  $\mu_0^*$ & $\widetilde{\mu}_0^{(A1,A2)}$& 0.02\% & 0.91\% & 1.08\% & 89.76\% & 1.08\%\\  
 & $\widetilde{\mu}_0^{(A1,A3)}$ & -0.05\% & 1.35\% & 1.35\% & 94.64\% &  1.35\%\\ 
 & $\widetilde{\mu}_0^{(A1,A2,A3)}$ & -0.02\% & 0.90\% & 1.01\% & 91.26\% &  1.01\%\\ 
  \hline
 \hline
$\nu_{1}^*$ & $\widetilde{\nu}_1^{(A1)}$ & 0.01\% & 1.39\% & 1.41\% & 94.44\% &  1.41\%\\ 
 &$\widetilde{\nu}_1^{(A1,A2,A3)}$& -0.03\% & 1.05\% & 1.01\% & 96.00\% &  1.01\%\\ 
   \hline
  $\nu_{0}^*$ & $\widetilde{\nu}_0^{(A1)}$ & -0.04\% & 1.32\% & 1.34\% & 94.38\% & 1.34\% \\ 
 & $\widetilde{\nu}_0^{(A1,A2,A3)}$& -0.02\% & 0.98\% & 1.08\% & 91.74\% &  1.08\%\\ 
  \hline    
  \hlinewd{1pt}\\[-1em]
  \multicolumn{7}{c}{(d) $\tau_t^*(X)=\tau_t^*(1,X)=\tau_t^*(0,X)$ modeled incorrectly, $\lambda_1^*(X)$ and $\pi_t^*(0,X)$ modeled correctly} \\ \hline
 Parameter&Estimator& {Bias} & Mean SE & SD & \parbox{2cm}{\centering 95\% CI \\ coverage} & $\sqrt{\mbox{MSE}}$ \\
  \hline
  $\mu_1^*$ & $\widetilde{\mu}_1^{(A1,A2)}$ & -0.20\% & 1.02\%& 1.00\%& 95.04\% & 1.02\% \\ 
 &$\widetilde{\mu}_1^{(A1,A3)}$ & -0.28\%& 1.60\% & 1.53\% & 94.74\% & 1.55\%\\ 
 &$\widetilde{\mu}_1^{(A1,A2,A3)}$ & -0.13\% & 1.00\% & 0.97\% & 95.54\% &  0.97\%\\ 
  \hline
  $\mu_0^*$ & $\widetilde{\mu}_0^{(A1,A2)}$& -0.55\% & 1.31\%& 1.21\% & 93.70\% & 1.32\% \\ 
 & $\widetilde{\mu}_0^{(A1,A3)}$ & -0.27\% & 1.54\% & 1.48\% & 95.26\% & 1.50\% \\ 
 & $\widetilde{\mu}_0^{(A1,A2,A3)}$ & -0.21\% & 1.15\% & 1.10\% & 95.40\% &  1.12\%\\ 
  \hline
 \hline
$\nu_{1}^*$ & $\widetilde{\nu}_1^{(A1)}$ & -0.04\%& 1.49\%& 1.51\%& 93.90\%& 1.51\% \\ 
 &$\widetilde{\nu}_1^{(A1,A2,A3)}$& -0.16\% & 1.18\%& 1.04\% & 96.92\% & 1.05\% \\ 
   \hline
  $\nu_{0}^*$ & $\widetilde{\nu}_0^{(A1)}$ & -0.03\% & 1.44\% & 1.47\% & 94.48\%& 1.47\% \\ 
 & $\widetilde{\nu}_0^{(A1,A2,A3)}$& -0.14\% & 1.24\% & 1.16\%& 96.28\%& 1.16\% \\ 
  \hline
   \hlinewd{1pt}
\end{tabular}
\ContinuedFloat
\end{table}

\begin{table}
\ContinuedFloat
\footnotesize
\centering
\caption{Simulation results: (A1), (A2), (A3) (continued)} 
\begin{tabular}{llccccc}
  \multicolumn{7}{c}{(e)  $\lambda_1^*(X)$, $\pi_t^*(0,X)$ modeled incorrectly, $\tau_t^*(X)=\tau_t^*(1,X)=\tau_t^*(0,X)$ modeled correctly} \\ \hline
 Parameter&Estimator& {Bias} & Mean SE & SD & \parbox{2cm}{\centering 95\% CI \\ coverage} & $\sqrt{\mbox{MSE}}$ \\
  \hline
  $\mu_1^*$ & $\widetilde{\mu}_1^{(A1,A2)}$ & 0.02\% & 0.97\% & 0.93\% & 95.50\% & 0.93\% \\ 
 &$\widetilde{\mu}_1^{(A1,A3)}$ & 0.04\% & 1.86\% & 1.41\% & 99.06\% & 1.41\% \\ 
 &$\widetilde{\mu}_1^{(A1,A2,A3)}$ & 0.02 & 0.95\% & 0.90\% & 95.66\% &  0.91\%\\ 
  \hline
  $\mu_0^*$ & $\widetilde{\mu}_0^{(A1,A2)}$& 0.03\% & 0.90\% & 1.06\% & 90.60\% & 1.06\% \\ 
 & $\widetilde{\mu}_0^{(A1,A3)}$ & 0.02\% & 1.77\% & 1.36\% & 98.54\% & 1.36\% \\ 
  & $\widetilde{\mu}_0^{(A1,A2,A3)}$ & 0.02\% & 0.90\% & 1.00\% & 92.24\% &  1.00\%\\ 
 \hline
 \hline
$\nu_{1}^*$ & $\widetilde{\nu}_1^{(A1)}$ & 0.03\% & 1.37\% & 1.39\% & 94.30\% & 1.39\% \\ 
 &$\widetilde{\nu}_1^{(A1,A2,A3)}$& 0.01\% & 0.78\% & 0.98\% & 88.16\% & 0.98\% \\ 
   \hline
  $\nu_{0}^*$ & $\widetilde{\nu}_0^{(A1)}$ & 0.02\% & 1.31\% & 1.35\% & 93.90\% & 1.35\% \\ 
 & $\widetilde{\nu}_0^{(A1,A2,A3)}$& -0.00\% & 0.75\% & 1.07\% & 82.84\% & 1.07\% \\ 
  \hline
      \multicolumn{7}{c}{(f) $\lambda_1^*(X)$,  $\tau_t^*(X)=\tau_t^*(1,X)=\tau_t^*(0,X)$ modeled incorrectly, $\pi_t^*(0,X)$ modeled correctly,} \\ \hline
 Parameter&Estimator& {Bias} & Mean SE & SD & \parbox{2cm}{\centering 95\% CI \\ coverage} & $\sqrt{\mbox{MSE}}$ \\
  \hline
  $\mu_1^*$ & $\widetilde{\mu}_1^{(A1,A2)}$  & -0.17\% & 1.03\%& 1.00\%& 95.02\% & 1.01\% \\ 
 &$\widetilde{\mu}_1^{(A1,A3)}$ & -7.17\%& 2.14\%& 1.67\%& 5.40\% & 7.37\% \\ 
 &$\widetilde{\mu}_1^{(A1,A2,A3)}$ & 0.46\% & 1.00\% & 0.98\% & 93.44\% &  1.83\%\\ 
  \hline
  $\mu_0^*$ & $\widetilde{\mu}_0^{(A1,A2)}$& -0.53\%& 1.32\%& 1.21\% & 94.40\% & 1.32\% \\ 
 & $\widetilde{\mu}_0^{(A1,A3)}$ & -7.11\% & 2.06\% & 1.63\% & 3.98\% & 7.29\%\\ 
  & $\widetilde{\mu}_0^{(A1,A2,A3)}$ & -1.45\% & 1.24\% & 1.12\% & 79.92\% &  1.83\%\\ 
  \hline
 \hline
$\nu_{1}^*$ & $\widetilde{\nu}_1^{(A1)}$ & -0.01\% & 1.50\% & 1.48\% & 95.32\% & 1.48\% \\ 
 &$\widetilde{\nu}_1^{(A1,A2,A3)}$& 5.78\% & 0.86\% & 1.02\% & 0.00\% & 5.86\% \\ 
   \hline
  $\nu_{0}^*$ & $\widetilde{\nu}_0^{(A1)}$ & -0.01\% & 1.45\% & 1.45\% & 94.84\% & 1.45\% \\  
  & $\widetilde{\nu}_0^{(A1,A2,A3)}$& 4.31\% & 1.00\% & 1.13\% & 1.32\% & 4.46\% \\ 
  \hline
      \hlinewd{1pt}
     \multicolumn{7}{c}{(g) $\pi_t^*(0,X)$, $\tau_t^*(X)=\tau_t^*(1,X)=\tau_t^*(0,X)$  modeled incorrectly, $\lambda_1^*(X)$ modeled correctly} \\ \hline
  Parameter&Estimator& {Bias} & Mean SE & SD & \parbox{2cm}{\centering 95\% CI \\ coverage} & $\sqrt{\mbox{MSE}}$ \\
  \hline
  $\mu_1^*$ & $\widetilde{\mu}_1^{(A1,A2)}$ & -1.67\% & 1.06\%& 0.99\% & 64.44\%& 1.94\% \\ 
 &$\widetilde{\mu}_1^{(A1,A3)}$ & -0.26\%& 1.61\%& 1.55\% & 94.98\%& 1.57\%\\ 
  &$\widetilde{\mu}_1^{(A1,A2,A3)}$ & -1.86\% & 1.06\% & 0.98\% & 58.50\% &  2.10\%\\ 
 \hline
  $\mu_0^*$ & $\widetilde{\mu}_0^{(A1,A2)}$& 2.04\%& 0.99\%& 1.17\% & 46.38\% & 2.35\%\\ 
 & $\widetilde{\mu}_0^{(A1,A3)}$ & -0.27\%& 1.54\% & 1.49\% & 95.34\% & 1.51\% \\ 
  & $\widetilde{\mu}_0^{(A1,A2,A3)}$ & 2.13\% & 1.01\% & 1.08\% & 44.86\% &  2.39\%\\ 
  \hline
 \hline
$\nu_{1}^*$ & $\widetilde{\nu}_1^{(A1)}$ & -0.02\% & 1.50\% & 1.52\% & 94.54\% & 1.52\% \\ 
 &$\widetilde{\nu}_1^{(A1,A2,A3)}$& -1.75\%& 1.25\% & 1.06\%& 73.02\%& 2.05\% \\ 
   \hline
  $\nu_{0}^*$ & $\widetilde{\nu}_0^{(A1)}$  & -0.01\%& 1.44\%& 1.47\% & 94.62\% & 1.47\% \\ 
 & $\widetilde{\nu}_0^{(A1,A2,A3)}$& 1.99\% & 1.12\% & 1.13\% & 57.58\% & 2.29\% \\ 
  \hline
   \hlinewd{1pt}
       \multicolumn{7}{c}{(h) $\lambda_1^*(X)$, $\pi_t^*(0,X)$, $\tau_t^*(X)=\tau_t^*(1,X)=\tau_t^*(0,X)$  modeled incorrectly} \\ \hline
 Parameter&Estimator& {Bias} & Mean SE & SD & \parbox{2cm}{\centering 95\% CI \\ coverage} & $\sqrt{\mbox{MSE}}$ \\
  \hline
  $\mu_1^*$ & $\widetilde{\mu}_1^{(A1,A2)}$ & -1.66\% & 1.08\% & 1.01\% & 66.22\% & 1.94\% \\ 
 &$\widetilde{\mu}_1^{(A1,A3)}$ & -7.25\% & 2.16\% & 1.74\% & 5.28\% & 7.46\% \\ 
   &$\widetilde{\mu}_1^{(A1,A2,A3)}$ & -1.56\% & 1.07\% & 1.00\% & 69.68\% &  1.85\%\\ 
 \hline
  $\mu_0^*$ & $\widetilde{\mu}_0^{(A1,A2)}$ & 2.10\% & 1.02\% & 1.22\% & 46.32\% & 2.43\% \\
 & $\widetilde{\mu}_0^{(A1,A3)}$ & -7.21\% & 2.07\% & 1.65\% & 4.10\% & 7.40\% \\ 
 & $\widetilde{\nu}_0^{(A1,A2,A3)}$& 1.86\% & 1.04\% & 1.11\% & 58.04\% & 2.16\% \\ 
  \hline
 \hline
$\nu_{1}^*$ & $\widetilde{\nu}_1^{(A1)}$ & -0.02\% & 1.53\% & 1.53\% & 95.10\% & 1.53\% \\
 &$\widetilde{\nu}_1^{(A1,A2,A3)}$& 4.38\% & 0.93\% & 1.03\% & 0.52\% & 4.50\% \\ 
   \hline
  $\nu_{0}^*$ & $\widetilde{\nu}_0^{(A1)}$ & -0.01\% & 1.47\% & 1.49\% & 94.12\% & 1.49\% \\ 
 & $\widetilde{\nu}_0^{(A1,A2,A3)}$& 6.70\% & 0.87\% & 1.14\% & 0.00\% & 6.79\% \\ 
  \hline
   \hlinewd{1pt}
\end{tabular}
\end{table}

\newpage

\section{R Code}

We refer to the data frame for the dataset as \texttt{datc}, let \texttt{n} denote the sample size and let \texttt{K} denote the number of sample splits.  We store the sample split results for each of the five estimators in matrices named \texttt{resmua1a2},  \texttt{resmua1a3}, \texttt{resmua1a2a3}, \texttt{resnua1} and \texttt{resnua1a2a3}, respectively.  Here is sample R code.

In the R code, the covariates have the following naming conventions: age (\texttt{AGE}), sex (\texttt{SEX}), highest level of education (\texttt{EDUC}), systolic blood pressure (\texttt{SBP}), diastolic blood pressure (\texttt{DBP}), qualifying symptoms (unstable angina/MI vs. other; \texttt{ANGTYPE}), number of diseased vessels (three vs. less than three; \texttt{DISREGB}), proximal left anterior descending disease (\texttt{TSPLADD}), prior myocardial infarction (\texttt{MI}), diabetes (no, with treatment, without treatment; \texttt{DIAB}),  current smoking (\texttt{CIG}), hypertension (\texttt{HYPER}).   In addition, \texttt{R} denotes the randomization consent indicator, \texttt{PTCA} denotes the indicator of receiving PTCA 
and \texttt{DEATH} denotes the binary indicator of death by the end of 5 years (\texttt{DEATH}). 

 \begin{singlespace}
{\tiny
\begin{verbatim}
K=5	# number of sample splits

resmua1a2 = matrix(0,K,4)
resmua1a3 = matrix(0,K,4)
resmua1a2a3 = matrix(0,K,4)
resnua1 = matrix(0,K,6)
resnua1a2a3 = matrix(0,K,6)

sam = sample(1:K,n,replace=TRUE)

for (k in 1: K) {
  
  datmk = datc[sam!=k,]	# minus kth split
  datk = datc[sam==k,] 	# kth split
  
  # gam model for death for PTCA patients
  # predicted probabilties on minus kth split and kth split
  gamyres1mk = gam(DEATH ~ SEX + DIAB + MI  + HYPER + TSPLADD + CIGB + 
                     DISREGB + EDUC + ANGTYPE + s(SBPB) +s(DBPB) + s(AGE), 
                   family=binomial, data = datmk, subset = (PTCA==1))   
  pgamyres1mk = predict.Gam(gamyres1mk,datmk,type='response')
  pgamyres1k = predict.Gam(gamyres1mk,datk,type='response')

  # gam model for death for CABG patients
  # predicted probabilties on minus kth split and kth split
  gamyres0mk = gam(DEATH ~ SEX + DIAB + MI  + HYPER + TSPLADD + CIGB + 
                     DISREGB + EDUC + ANGTYPE + s(SBPB) +s(DBPB) + s(AGE), 
                   family=binomial, data = datmk, subset = (PTCA==0))   
  pgamyres0mk = predict.Gam(gamyres0mk,datmk,type='response')
  pgamyres0k = predict.Gam(gamyres0mk,datk,type='response')
  
  # gam model for death for PTCA patients in RCT
  # predicted probabilties on minus kth split and kth split
  gamyres11mk = gam(DEATH ~ SEX + DIAB + MI  + HYPER + TSPLADD + CIGB + 
                      DISREGB + EDUC + ANGTYPE + s(SBPB) +s(DBPB) + s(AGE), 
                    family=binomial, data = datmk, subset = (R==1 & PTCA==1))   
  pgamyres11mk = predict.Gam(gamyres11mk,datmk,type='response')
  pgamyres11k = predict.Gam(gamyres11mk,datk,type='response')
  
  # gam model for death for CABG patients in RCT
  # predicted probabilties on minus kth split and kth split
  gamyres10mk = gam(DEATH ~ SEX + DIAB + MI  + HYPER + TSPLADD + CIGB + 
                      DISREGB + EDUC + ANGTYPE + s(SBPB) +s(DBPB) + s(AGE), 
                    family=binomial, data = datmk, subset = (R==1 & PTCA==0))   
  pgamyres10mk = predict.Gam(gamyres10mk,datmk,type='response')
  pgamyres10k = predict.Gam(gamyres10mk,datk,type='response')
  
  # gam model for death for PTCA patients in OBS
  # predicted probabilties on minus kth split and kth split
  gamyres01mk = gam(DEATH ~ SEX + DIAB + MI  + HYPER + TSPLADD + CIGB + 
                      DISREGB + EDUC + ANGTYPE + s(SBPB) +s(DBPB) + s(AGE), 
                    family=binomial, data = datmk, subset = (R==0 & PTCA==1))   
  pgamyres01mk = predict.Gam(gamyres01mk,datmk,type='response')
  pgamyres01k = predict.Gam(gamyres01mk,datk,type='response')
  
  # gam model for death for CABG patients in OBS
  # predicted probabilties on minus kth split and kth split
  gamyres00mk = gam(DEATH ~ SEX + DIAB + MI  + HYPER + TSPLADD + CIGB + 
                      DISREGB + EDUC + ANGTYPE + s(SBPB) +s(DBPB) + s(AGE), 
                    family=binomial, data = datmk, subset = (R==0 & PTCA==0))   
  pgamyres00mk = predict.Gam(gamyres00mk,datmk,type='response')
  pgamyres00k = predict.Gam(gamyres00mk,datk,type='response')
  
  # gam model for enrollment into RCT
  # predicted probabilties on minus kth split and kth split
  gamrresmk = gam(R ~ SEX + DIAB + MI  + HYPER + TSPLADD + CIGB + 
                    DISREGB + EDUC + ANGTYPE + s(SBPB) +s(DBPB) + s(AGE), 
                  family=binomial, data = datmk)   
  pgamrresmk = predict.Gam(gamrresmk,datmk,type='response')
  pgamrresk = predict.Gam(gamrresmk,datk,type='response')
  
  # gam model for PTCA in OBS
  # predicted probabilties on minus kth split and kth split
  gamtres0mk = gam(PTCA ~ SEX + DIAB + MI  + HYPER + TSPLADD + CIGB + 
                     DISREGB + EDUC + ANGTYPE + s(SBPB) +s(DBPB) + s(AGE), 
                   family=binomial, data = datmk, subset = (R==0))   
  pgamtres0mk = predict.Gam(gamtres0mk,datmk,type='response')
  pgamtres0k = predict.Gam(gamtres0mk,datk,type='response')
   
  # estimated conditional probability of death under PTCA (CABG) given study and covariates
  # estimated on minus kth split and kth split
  tau1rxmk = pgamyres11mk * (datmk$R==1) + pgamyres01mk * (datmk$R==0)
  tau0rxmk = pgamyres10mk * (datmk$R==1) + pgamyres00mk * (datmk$R==0)
  tau1rxk = pgamyres11k * (datk$R==1) + pgamyres01k * (datk$R==0)
  tau0rxk = pgamyres10k * (datk$R==1) + pgamyres00k * (datk$R==0)
 
  # estimated conditional probability of death under PTCA (CABG) given covariates
  # estimated on minus kth split and kth split 
  tau1xmk = pgamyres1mk 
  tau0xmk = pgamyres0mk 
  tau1xk =  pgamyres1k 
  tau0xk = pgamyres0k
  
  # probability of PTCA in RCT for minus kth split and kth split
  pi11mk = 0.5
  pi11k = 0.5
  
  # marginal probability of study type for minus kth split and kth split
  lambda1mk = mean(datmk$R)
  lambda1k = mean(datk$R)
 
  # estimated conditional probability of PTCA (CABG) given study and  covariates
  # estimated on minus kth split and kth split  
  pi1rxmk = pgamtres0mk * (datmk$R==0) + pi11mk * (datmk$R==1)
  pi0rxmk = (1-pgamtres0mk) * (datmk$R==0) + (1-pi11mk) * (datmk$R==1)
  pi1rxk = pgamtres0k * (datk$R==0) + pi11mk * (datk$R==1)
  pi0rxk = (1-pgamtres0k) * (datk$R==0) + (1-pi11mk) * (datk$R==1)
 
  # estimated conditional probability of PTCA (CABG) given covariates
  # estimated on minus kth split and kth split   
  pi1xmk = pgamrresmk * pi11mk + (1-pgamrresmk) * pgamtres0mk
  pi0xmk = pgamrresmk * (1-pi11mk) + (1-pgamrresmk) * (1-pgamtres0mk)
  pi1xk = pgamrresk * pi11mk + (1-pgamrresk) * pgamtres0k
  pi0xk = pgamrresk * (1-pi11mk) + (1-pgamrresk) * (1-pgamtres0k)
  
  # kth contribution to estimate of mu under Assumptions A1, A2
  # kth contribution to standard error calculation
  resmua1a2[k,1] = mean((datk$PTCA==1)*(datk$DEATH)/pi1rxk + 
                          (1-(datk$PTCA==1)/pi1rxk)*tau1rxk)
  resmua1a2[k,2] = mean((datk$PTCA==0)*(datk$DEATH)/pi0rxk + 
                          (1-(datk$PTCA==0)/pi0rxk)*tau0rxk) 
  resmua1a2[k,3] = sum(((datk$PTCA==1)*(datk$DEATH)/pi1rxk + 
                          (1-(datk$PTCA==1)/pi1rxk)*tau1rxk-resmua1a2[k,1])^2) 
  resmua1a2[k,4] = sum(((datk$PTCA==0)*(datk$DEATH)/pi0rxk + 
                          (1-(datk$PTCA==0)/pi0rxk)*tau0rxk-resmua1a2[k,2])^2) 
  
  # kth contribution to estimate of mu under Assumptions A1, A3
  # kth contribution to standard error calculation
  resmua1a3[k,1] = mean((datk$PTCA==1)*(datk$R)*(datk$DEATH)/(pgamrresk*pi11mk) + 
                          (1-(datk$PTCA==1)*(datk$R)/(pgamrresk*pi11mk))*pgamyres11k)
  resmua1a3[k,2] = mean((datk$PTCA==0)*(datk$R)*(datk$DEATH)/(pgamrresk*(1-pi11mk)) + 
                          (1-(datk$PTCA==0)*(datk$R)/(pgamrresk*(1-pi11mk)))*pgamyres10k)
  resmua1a3[k,3] = sum(((datk$PTCA==1)*(datk$R)*(datk$DEATH)/(pgamrresk*pi11mk) + 
                          (1-(datk$PTCA==1)*(datk$R)/(pgamrresk*pi11mk))*pgamyres11k-resmua1a3[k,1])^2)
  resmua1a3[k,4] = sum(((datk$PTCA==0)*(datk$R)*(datk$DEATH)/(pgamrresk*(1-pi11mk)) + 
                          (1-(datk$PTCA==0)*(datk$R)/(pgamrresk*(1-pi11mk)))*pgamyres10k-resmua1a3[k,2])^2)
  
  # kth contribution to estimate of mu under Assumptions A1, A2, A3
  # kth contribution to standard error calculation
  resmua1a2a3[k,1] = mean((datk$PTCA==1)*(datk$DEATH)/pi1xk +
                            (1-(datk$PTCA==1)/pi1xk)*tau1xk) 
  resmua1a2a3[k,2] = mean((datk$PTCA==0)*(datk$DEATH)/pi0xk + 
                            (1-(datk$PTCA==0)/pi0xk)*tau0xk) 
  resmua1a2a3[k,3] = sum(((datk$PTCA==1)*(datk$DEATH)/pi1xk + 
                            (1-(datk$PTCA==1)/pi1xk)*tau1xk-resmua1a2a3[k,1])^2) 
  resmua1a2a3[k,4] = sum(((datk$PTCA==0)*(datk$DEATH)/pi0xk + 
                            (1-(datk$PTCA==0)/pi0xk)*tau0xk-resmua1a2a3[k,2])^2) 
  
  # kth contribution to estimate of nu under Assumption A1
  # kth contribution to standard error calculation
  resnua1[k,1] = mean(datmk$R*pgamyres11mk/lambda1mk)  
  resnua1[k,2] = mean(datmk$R*pgamyres10mk/lambda1mk)  
  resnua1[k,3] = resnua1[k,1] + 
    mean((datk$PTCA==1)*(datk$R)*(datk$DEATH)/(lambda1mk*pi11mk) + 
           datk$R/lambda1mk * (1-(datk$PTCA==1)/pi11mk)*pgamyres11k - 
           datk$R/lambda1mk*resnua1[k,1])
  resnua1[k,4] = resnua1[k,2] + 
    mean((datk$PTCA==0)*(datk$R)*(datk$DEATH)/(lambda1mk*(1-pi11mk)) + 
           datk$R/lambda1mk * (1-(datk$PTCA==0)/(1-pi11mk))*pgamyres10k - 
           datk$R/lambda1mk*resnua1[k,2])
  resnua1[k,5] = sum(((datk$PTCA==1)*(datk$R)*(datk$DEATH)/(lambda1mk*pi11mk) + 
                        datk$R/lambda1mk * (1-(datk$PTCA==1)/pi11mk)*pgamyres11k - 
                        datk$R/lambda1mk*resnua1[k,3])^2)
  resnua1[k,6] = sum(((datk$PTCA==0)*(datk$R)*(datk$DEATH)/(lambda1mk*(1-pi11mk)) +
                        datk$R/lambda1mk * (1-(datk$PTCA==0)/(1-pi11mk))*pgamyres10k - 
                        datk$R/lambda1mk*resnua1[k,4])^2)
  
  # kth contribution to estimate of nu under Assumptions A1, A2, A3
  # kth contribution to standard error calculation
  resnua1a2a3[k,1] = mean(pgamrresmk*tau1xmk/lambda1mk)  
  resnua1a2a3[k,2] = mean(pgamrresmk*tau0xmk/lambda1mk)  
  resnua1a2a3[k,3] = resnua1a2a3[k,1] + 
    mean((datk$PTCA==1)*pgamrresk*(datk$DEATH)/(lambda1mk*pi1xk) + 
           (datk$R-(datk$PTCA==1)*pgamrresk/pi1xk)*tau1xk/lambda1mk - 
           datk$R/lambda1mk*resnua1a2a3[k,1])
  resnua1a2a3[k,4] = resnua1a2a3[k,2] + 
    mean((datk$PTCA==0)*pgamrresk*(datk$DEATH)/(lambda1mk*(1-pi1xk)) + 
           (datk$R-(datk$PTCA==0)*pgamrresk/(1-pi1xk))*tau0xk/lambda1mk - 
           datk$R/lambda1mk*resnua1a2a3[k,2])
  resnua1a2a3[k,5] = sum(((datk$PTCA==1)*pgamrresk*(datk$DEATH)/(lambda1mk*pi1xk) + 
                            (datk$R-(datk$PTCA==1)*pgamrresk/pi1xk)*tau1xk/lambda1mk - 
                            datk$R/lambda1mk*resnua1a2a3[k,3])^2)
  resnua1a2a3[k,6] = sum(((datk$PTCA==0)*pgamrresk*(datk$DEATH)/(lambda1mk*(1-pi1xk)) + 
                            (datk$R-(datk$PTCA==0)*pgamrresk/(1-pi1xk))*tau0xk/lambda1mk - 
                            datk$R/lambda1mk*resnua1a2a3[k,4])^2)
}

# Estimates, standard errors and confidence intervals of mu, nu
mua1a2 = apply(resmua1a2,2,mean)[1:2]
semua1a2 = sqrt(apply(resmua1a2,2,sum)/n^2)[3:4]
lcimua1a2 = mua1a2 - 1.96*semua1a2
ucimua1a2 = mua1a2 + 1.96*semua1a2
mua1a3 = apply(resmua1a3,2,mean)[1:2]
semua1a3 = sqrt(apply(resmua1a3,2,sum)/n^2)[3:4]
lcimua1a3 = mua1a3 - 1.96*semua1a3
ucimua1a3 = mua1a3 + 1.96*semua1a3
mua1a2a3 = apply(resmua1a2a3,2,mean)[1:2]
semua1a2a3 = sqrt(apply(resmua1a2a3,2,sum)/n^2)[3:4]
lcimua1a2a3 = mua1a2a3 - 1.96*semua1a2a3
ucimua1a2a3 = mua1a2a3+ 1.96*semua1a2a3
nua1 = apply(resnua1,2,mean)[3:4]
senua1 = sqrt(apply(resnua1,2,sum)/n^2)[5:6]
lcinua1 = nua1 - 1.96*senua1
ucinua1 = nua1 + 1.96*senua1
nua1a2a3 = apply(resnua1a2a3,2,mean)[3:4]
senua1a2a3 = sqrt(apply(resnua1a2a3,2,sum)/n^2)[5:6]
lcinua1a2a3 = nua1a2a3 - 1.96*senua1a2a3
ucinua1a2a3 = nua1a2a3 + 1.96*senua1a2a3

# Estimates, standard errors and confidence intervals of treatment effects
deltamua1a2 = mua1a2[1]-mua1a2[2]
sedeltamua1a2 = sqrt(sum(semua1a2^2))
cideltamua1a2 =c(deltamua1a2-1.96*sedeltamua1a2,
                 deltamua1a2+1.96*sedeltamua1a2)
deltamua1a3 = mua1a3[1]-mua1a3[2]
sedeltamua1a3 = sqrt(sum(semua1a3^2))
cideltamua1a3 =c(deltamua1a3-1.96*sedeltamua1a3,
                 deltamua1a3+1.96*sedeltamua1a3)
deltamua1a2a3 = mua1a2a3[1]-mua1a2a3[2]
sedeltamua1a2a3 = sqrt(sum(semua1a2a3^2))
cideltamua1a2a3 =c(deltamua1a2a3-1.96*sedeltamua1a2a3,
                   deltamua1a2a3+1.96*sedeltamua1a2a3)
deltanua1 = nua1[1]-nua1[2]
sedeltanua1 = sqrt(sum(senua1^2))
cideltanua1 =c(deltanua1-1.96*sedeltanua1,
               deltanua1+1.96*sedeltanua1)
deltanua1a2a3 = nua1a2a3[1]-nua1a2a3[2]
sedeltanua1a2a3 = sqrt(sum(senua1a2a3^2))
cideltanua1a2a3 =c(deltanua1a2a3-1.96*sedeltanua1a2a3,
                   deltanua1a2a3+1.96*sedeltanua1a2a3)
\end{verbatim}
}

\end{singlespace}

\end{appendices}

\end{document}